\documentclass[journal, romanappendices]{IEEEtran}
\usepackage{paralist}
\usepackage{graphicx}
\usepackage{amsmath}
\usepackage{balance}
\usepackage{cite}
\usepackage{overpic}
\usepackage{url}
\usepackage{booktabs} 
\usepackage{vmr-symbols-vecbold}
\usepackage{standard-macros}
\PassOptionsToPackage{hyphens}{url}
\usepackage{xcolor}
\definecolor{links}{rgb}{0.5,0,0}   
\definecolor{urls}{rgb}{0,0,0.8}    
\definecolor{cites}{rgb}{0,0,0.6}   
\usepackage[colorlinks,hyperindex,linkcolor=links,citecolor=cites,urlcolor=urls]{hyperref} 

\DeclareSymbolFontAlphabet{\amsmathbb}{AMSb}%

\newcommand{\lro}[1]{\lefto({#1}\right)}																
\newcommand{\lrho}[1]{\lefto [ {#1} \right ]}																				

\newcommand{\lr}[1]{\left({#1}\right)}																
\newcommand{\lrb}[1]{\left \lbrace {#1} \right \rbrace}															

\safemath{\dopplerspread}{B_D}																								
\safemath{\delayspread}{T_D}																									
\safemath{\nc}{n\sub{c}}																										
\safemath{\nf}{n\sub{f}}																										
\safemath{\efa}{p\sub{sc}}
\safemath{\efb}{p\sub{cs}}
\safemath{\ef}{\epsilon\sub{f}	}
\safemath{\nd}{n\sub{d}}																										
\safemath{\ntx}{n\sub{t}} 																											
\safemath{\nrx}{n\sub{r}}																											
\safemath{\ntxt}{\tilde{n\sub{t}}}																											
\safemath{\cb}{\ensuremath{L}} 																								
\safemath{\cl}{\ensuremath{n}} 																								
\safemath{\txanto}{{\ensuremath{\tilde{m}_t}}} 																		
\safemath{\cs}{M} 																														
\safemath{\idPustm}{\ensuremath{S_{k}}}
\safemath{\error}{\ensuremath{\epsilon}} 																				
\safemath{\eexp}{\ensuremath{\mathcal{E}}} 																			
\safemath{\nsubc}{n\sub{s}}			 																						
\safemath{\nofdm}{n\sub{o}} 																									
\safemath{\bc}{\ensuremath{B_c}} 																							
\safemath{\ts}{\ensuremath{T_s}} 																							
\safemath{\nrb}{\ensuremath{n_{rb}}} 																						
\safemath{\rul}{\ensuremath{\rho\sub{ul}}}
\safemath{\rdl}{\ensuremath{\rho\sub{dl}}}

\safemath{\nres}{\ell}
\safemath{\nr}{n\sub{r}}
\safemath{\maxk}{M^*\lr{\nres, \nsubc, \nofdm, \epsilon, \rho}}
\safemath{\Rmax}{R^*}
\safemath{\Emin}{E\sub{b}^*/N_0}
\safemath{\Eminf}{\frac{E\sub{b}^*}{N_0}}
\safemath{\np}{\ensuremath{n\sub{p}}}
\safemath{\ndf}{\ensuremath{\bar{n}\sub{d}}}
\safemath{\npf}{\ensuremath{\bar{n}\sub{p}}}
\safemath{\code}{\ensuremath{\mathcal{C}}}
\safemath{\err}{\ensuremath{\epsilon}}
\safemath{\rp}{\ensuremath{\rho\sub{p}}}
\safemath{\rd}{\ensuremath{\rho\sub{d}}}
\safemath{\cohtime}{\ensuremath{T\sub{c}}}
\safemath{\cohbw}{\ensuremath{B\sub{c}}}
\safemath{\nmax}{\ensuremath{\ell\sub{m}}}
\safemath{\ntot}{\ensuremath{n\sub{tot}}}
\safemath{\nul}{\ensuremath{n^{\mathrm{ul}}}}
\safemath{\ndl}{\ensuremath{n^{\mathrm{dl}}}}

\safemath{\yp}{\ensuremath{\randvecy_{\nu}^{(\text{p})}}}
\safemath{\yd}{\ensuremath{\randvecy_{\nu}^{(\text{d})}}}
\safemath{\ypd}{\ensuremath{\vecy_{\nu}^{(\text{p})}}}
\safemath{\ydd}{\ensuremath{\vecy_{\nu}^{(\text{d})}}}

\safemath{\ypf}{\ensuremath{\bar{\randvecy}_{\nu}^{(\text{p})}}}
\safemath{\ydf}{\ensuremath{\bar{\randvecy}_{\nu}^{(\text{d})}}}
\safemath{\ypdf}{\ensuremath{\bar{\vecy}_{\nu}^{(\text{p})}}}
\safemath{\yddf}{\ensuremath{\bar{\vecy}_{\nu}^{(\text{d})}}}

\safemath{\xp}{\ensuremath{\vecx^{(\text{p})}}}
\safemath{\xd}{\ensuremath{\randvecx_{\nu}^{(\text{d})}}}
\safemath{\xdd}{\ensuremath{\vecx_{\nu}^{(\text{d})}}}

\safemath{\xpf}{\ensuremath{\bar{\vecx}^{(\text{p})}}}
\safemath{\xdf}{\ensuremath{\bar{\randvecx}_{\nu}^{(\text{d})}}}
\safemath{\xddf}{\ensuremath{\bar{\vecx}_{\nu}^{(\text{d})}}}

\safemath{\xdb}{\ensuremath{\overline{\randvecx}^{(\text{d})}}}
\safemath{\Pxd}{\ensuremath{P_{\randvecx^{(\text{d})}}}}

\safemath{\xpbar}{\ensuremath{\overline{\matX}^{(\text{p})}}}
\safemath{\xdbar}{\ensuremath{\overline{\randmatX}^{(\text{d})}}}

\safemath{\xdv}{\ensuremath{\randvecx^{(\text{d})}}}
\safemath{\xdbarv}{\ensuremath{\overline{\randvecx}^{(\text{d})}}}
\safemath{\ydv}{\ensuremath{\randvecy^{(\text{d})}}}

\safemath{\xdr}{\ensuremath{\matX^{(\text{d})}}}

\safemath{\ttx}{\ensuremath{\tau\sub{tx}}}
\safemath{\trx}{\ensuremath{\tau\sub{rx}}}
\safemath{\ack}{\ensuremath{\mathrm{s}}}
\safemath{\nack}{\ensuremath{\mathrm{c}}}

\newcommand{\prob}[1]{\ensuremath{\mathbb{P}\lrho{#1}}}

\safemath{\mI}{\ensuremath{i\lro{\randvecy ; \randvecx}}} 				

\safemath{\randveca}{\bm{A}}
\safemath{\randvecb}{\bm{B}}
\safemath{\randvecc}{\bm{C}}
\safemath{\randvecd}{\bm{D}}
\safemath{\randvece}{\bm{E}}
\safemath{\randvecf}{\bm{F}}
\safemath{\randvecg}{\bm{G}}
\safemath{\randvech}{\bm{H}}
\safemath{\randveci}{\bm{I}}
\safemath{\randvecj}{\bm{J}}
\safemath{\randveck}{\bm{K}}
\safemath{\randvecl}{\bm{L}}
\safemath{\randvecm}{\bm{M}}
\safemath{\randvecn}{\bm{N}}
\safemath{\randveco}{\bm{O}}
\safemath{\randvecp}{\bm{P}}
\safemath{\randvecq}{\bm{Q}}
\safemath{\randvecr}{\bm{R}}
\safemath{\randvecs}{\bm{S}}
\safemath{\randvect}{\bm{T}}
\safemath{\randvecu}{\bm{U}}
\safemath{\randvecv}{\bm{V}}
\safemath{\randvecw}{\bm{W}}
\safemath{\randvecx}{\bm{X}}
\safemath{\randvecy}{\bm{Y}}
\safemath{\randvecz}{\bm{Z}}
\safemath{\randvecphi}{\bm{\Phi}}

\safemath{\randmatA}{\amsmathbb{A}}
\safemath{\randmatB}{\amsmathbb{B}}
\safemath{\randmatC}{\amsmathbb{C}}
\safemath{\randmatD}{\amsmathbb{D}}
\safemath{\randmatE}{\amsmathbb{E}}
\safemath{\randmatF}{\amsmathbb{F}}
\safemath{\randmatG}{\amsmathbb{G}}
\safemath{\randmatH}{\amsmathbb{H}}
\safemath{\randmatI}{\amsmathbb{I}}
\safemath{\randmatJ}{\amsmathbb{J}}
\safemath{\randmatK}{\amsmathbb{K}}
\safemath{\randmatL}{\amsmathbb{L}}
\safemath{\randmatM}{\amsmathbb{M}}
\safemath{\randmatN}{\amsmathbb{N}}
\safemath{\randmatO}{\amsmathbb{O}}
\safemath{\randmatP}{\amsmathbb{P}}
\safemath{\randmatQ}{\amsmathbb{Q}}
\safemath{\randmatR}{\amsmathbb{R}}
\safemath{\randmatS}{\amsmathbb{S}}
\safemath{\randmatT}{\amsmathbb{T}}
\safemath{\randmatU}{\amsmathbb{U}}
\safemath{\randmatV}{\amsmathbb{V}}
\safemath{\randmatW}{\amsmathbb{W}}
\safemath{\randmatX}{\amsmathbb{X}}
\safemath{\randmatY}{\amsmathbb{Y}}
\safemath{\randmatZ}{\amsmathbb{Z}}
\safemath{\randmatSigma}{\mathbb{\Sigma}}
\safemath{\randmatPhi}{\mathbb{\Phi}}
\safemath{\randmatLambda}{\mathbb{\Lambda}}

\safemath{\matSigma}{\bm{\Sigma}}
\safemath{\matPhi}{\bm{\Phi}}
\safemath{\matLambda}{\bm{\Lambda}}

\usepackage{glossaries}
\glsdisablehyper
\newglossaryentry{snr}
{
  name={SNR},
  description={signal-to-noise ratio},
  first={\glsentrydesc{snr} (\glsentrytext{snr})}
}

\newglossaryentry{simo}
{
  name={SIMO},
  description={single-input multiple-output},
  first={\glsentrydesc{simo} (\glsentrytext{simo})}
}

\newglossaryentry{mmse}
{
  name={MMSE},
  description={Minimum mean-square error},
  first={\glsentrydesc{mmse} (\glsentrytext{mmse})}
}

\newglossaryentry{na}
{
  name={NA},
  description={normal approximation},
  first={\glsentrydesc{na} (\glsentrytext{na})},
  plural={NAs},
  descriptionplural={normal approximations},
  firstplural={\glsentrydescplural{na} (\glsentryplural{na})}
}

\newglossaryentry{zf}
{
  name={ZF},
  description={zero-forcing},
  first={\glsentrydesc{zf} (\glsentrytext{zf})}
}

\newglossaryentry{rzf}
{
  name={RZF},
  description={regularized zero-forcing},
  first={\glsentrydesc{rzf} (\glsentrytext{rzf})}
}

\newglossaryentry{sir}
{
  name={SIR},
  description={signal-to-interference ratio},
  first={\glsentrydesc{sir} (\glsentrytext{sir})}
}

\newglossaryentry{mse}
{
  name={MSE},
  description={mean-square error},
  first={\glsentrydesc{mse} (\glsentrytext{mse})}
}
\newglossaryentry{mmmse}
{
  name={M-MMSE},
  description={multicell MMSE},
  first={\glsentrydesc{mmmse} (\glsentrytext{mmmse})}
}

\newglossaryentry{ls}
{
  name={LS},
  description={least-squares},
  first={\glsentrydesc{ls} (\glsentrytext{ls})}
}

\newglossaryentry{mr}
{
  name={MR},
  description={maximum ratio},
  first={\glsentrydesc{mr} (\glsentrytext{mr})}
}

\newglossaryentry{ue}
{
  name={UE},
  description={user equipment},
  first={\glsentrydesc{ue} (\glsentrytext{ue})},
  plural={UEs},
  descriptionplural={user equipments},
  firstplural={\glsentrydescplural{ue} (\glsentryplural{ue})}
}

\newglossaryentry{ap}
{
  name={AP},
  description={access point},
  first={\glsentrydesc{ap} (\glsentrytext{ap})},
  plural={APs},
  descriptionplural={access points},
  firstplural={\glsentrydescplural{ap} (\glsentryplural{ap})}
}

\newglossaryentry{cpu}
{
  name={CPU},
  description={central processing unit},
  first={\glsentrydesc{cpu} (\glsentrytext{cpu})},
  plural={CPUs},
  descriptionplural={central processing units},
  firstplural={\glsentrydescplural{cpu} (\glsentryplural{cpu})}
}

\newglossaryentry{bs}
{
  name={BS},
  description={base station},
  first={\glsentrydesc{bs} (\glsentrytext{bs})},
  plural={BSs},
  descriptionplural={base stations},
  firstplural={\glsentrydescplural{bs} (\glsentryplural{bs})}
}

\newglossaryentry{ostbc}
{
  name={OSTBC},
  description={orthogonal space-time block code},
  first={\glsentrydesc{ostbc} (\glsentrytext{ostbc})},
  plural={OSTBCs},
  descriptionplural={orthogonal space-time block codes},
  firstplural={\glsentrydescplural{ostbc} (\glsentryplural{ostbc})}
}

\newglossaryentry{biawgn}
{
  name={bi-AWGN},
  description={binary-input additive white Gaussian noise},
  first={\glsentrydesc{biawgn} (\glsentrytext{biawgn})}
}

\newglossaryentry{flnf}
{
  name={FLNF},
  description={fixed-length no-feedback},
  first={\glsentrydesc{flnf} (\glsentrytext{flnf})}
}

\newglossaryentry{crc}
{
  name={CRC},
  description={cyclic redundancy check},
  first={\glsentrydesc{crc} (\glsentrytext{crc})}
}

\newglossaryentry{bsc}
{
  name={BSC},
  description={binary symmetric channel},
  first={\glsentrydesc{bsc} (\glsentrytext{bsc})}
  plural={BSCs},
  descriptionplural={binary symmetric channels},
  firstplural={\glsentrydescplural{bsc} (\glsentryplural{bsc})}
}

\newglossaryentry{bec}
{
  name={BEC},
  description={binary-erasure channel},
  first={\glsentrydesc{bec} (\glsentrytext{bec})}
}
\newglossaryentry{awgn}
{
  name={AWGN},
  description={additive white Gaussian noise},
  first={\glsentrydesc{awgn} (\glsentrytext{awgn})}
}
\newglossaryentry{3gpp}
{
  name={3GPP},
  description={3rd generation partnership project},
  first={\glsentrydesc{3gpp} (\glsentrytext{3gpp})}
}

\newglossaryentry{dl}
{
  name={DL},
  description={downlink},
  first={\glsentrydesc{dl} (\glsentrytext{dl})}
}

\newglossaryentry{LED}
{
  name={LED},
  description={light emitting diode},
  first={\glsentrydesc{LED} (\glsentrytext{LED})},
  plural={LEDs},
  descriptionplural={light emitting diodes},
  firstplural={\glsentrydescplural{LED} (\glsentryplural{LED})}
}

\newglossaryentry{lte}
{
  name={LTE},
  description={long term evolution},
  first={\glsentrydesc{lte} (\glsentrytext{lte})}
}
\newglossaryentry{ofdm}
{
  name={OFDM},
  description={orthogonal frequency-division multiplexing},
  first={\glsentrydesc{ofdm} (\glsentrytext{ofdm})}
}

\newglossaryentry{ul}
{
  name={UL},
  description={uplink},
  first={\glsentrydesc{ul} (\glsentrytext{ul})}
}
\newglossaryentry{rb}
{
  name={RB},
  description={resource block},
  first={\glsentrydesc{rb} (\glsentrytext{rb})},
  plural={RBs},
  descriptionplural={resource blocks},
  firstplural={\glsentrydescplural{rb} (\glsentryplural{rb})}
}
\newglossaryentry{cu}
{
  name={CU},
  description={channel use},
  first={\glsentrydesc{cu} (\glsentrytext{cu})},
  plural={CUs},
  descriptionplural={channel uses},
  firstplural={\glsentrydescplural{cu} (\glsentryplural{cu})}
}
\newglossaryentry{tti}
{
  name={TTI},
  description={transmission time interval},
  first={\glsentrydesc{tti} (\glsentrytext{tti})},
  plural={TTI's},
  descriptionplural={transmission time intervals},
  firstplural={\glsentrydescplural{tti} (\glsentryplural{tti})}
}
\newglossaryentry{iid}
{
  name={i.i.d.},
  description={independent and identically distributed},
  first={\glsentrydesc{iid} (\glsentrytext{iid})}
}
\newglossaryentry{psd}
{
  name={PSD},
  description={power spectral density},
  first={\glsentrydesc{psd} (\glsentrytext{psd})}
}

\newglossaryentry{mimo}
{
  name={MIMO},
  description={multiple-input multiple-output},
  first={\glsentrydesc{mimo} (\glsentrytext{mimo})}
}

\newglossaryentry{bler}
{
  name={BLER},
  description={block error probability},
  first={\glsentrydesc{bler} (\glsentrytext{bler})}
}
\newglossaryentry{mtc}
{
  name={MTC},
  description={machine-type communication},
  first={\glsentrydesc{mtc} (\glsentrytext{mtc})}
}
\newacronym{csi}{CSI}{channel state information}
\newglossaryentry{dvbt}
{
  name={DVB-T},
  description={terrestrial digital video broadcasting},
  first={\glsentrydesc{dvbt} (\glsentrytext{dvbt})}
}
\newglossaryentry{pat}
{
  name={PAT},
  description={pilot-assisted transmission},
  first={\glsentrydesc{pat} (\glsentrytext{pat})}
}
\newglossaryentry{ml}
{
  name={ML},
  description={maximum likelihood},
  first={\glsentrydesc{ml} (\glsentrytext{ml})}
}
\newglossaryentry{dt}
{
  name={DT},
  description={dependence-testing},
  first={\glsentrydesc{dt} (\glsentrytext{dt})}
}
\newglossaryentry{ustm}
{
  name={USTM},
  description={unitary space-time modulation},
  first={\glsentrydesc{ustm} (\glsentrytext{ustm})}
}
\newglossaryentry{siso}
{
  name={SISO},
  description={single-input single-output},
  first={\glsentrydesc{siso} (\glsentrytext{siso})}
}
\newglossaryentry{snn}
{
  name={SNN},
  description={scaled nearest-neighbor},
  first={\glsentrydesc{snn} (\glsentrytext{snn})}
}

\newglossaryentry{snnd}
{
  name={SNND},
  description={scaled nearest-neighbor decoding},
  first={\glsentrydesc{snnd} (\glsentrytext{snnd})}
}
\newglossaryentry{rcus}
{
  name={RCUs},
  description={random-coding union bound with parameter $s$},
  first={\glsentrydesc{rcus} (\glsentrytext{rcus})}
}
\newglossaryentry{osd}
{
  name={OSD},
  description={ordered statistics decoding},
  first={\glsentrydesc{osd} (\glsentrytext{osd})}
}
\newglossaryentry{llr}
{
  name={LLR},
  description={log-likelihood ratio},
  first={\glsentrydesc{llr} (\glsentrytext{llr})}
   plural={LLR's},
  descriptionplural={log-likelihood ratios},
  firstplural={\glsentrydescplural{llr} (\glsentryplural{llr})}
}
\newglossaryentry{qpsk}
{
  name={QPSK},
  description={quaternary phase shift keying},
  first={\glsentrydesc{qpsk} (\glsentrytext{qpsk})}
}
\newglossaryentry{nlos}
{
  name={NLOS},
  description={non-line-of-sight},
  first={\glsentrydesc{nlos} (\glsentrytext{nlos})}
}
\newglossaryentry{los}
{
  name={LOS},
  description={line-of-sight},
  first={\glsentrydesc{los} (\glsentrytext{los})}
}
\newglossaryentry{mgf}
{
  name={MGF},
  description={moment-generating function},
  first={\glsentrydesc{mgf} (\glsentrytext{mgf})}
}
\newglossaryentry{cgf}
{
  name={CGF},
  description={cumulant-generating function},
  first={\glsentrydesc{cgf} (\glsentrytext{cgf})},
  plural={CGFs},
  descriptionplural={cumulant-generating functions},
  firstplural={\glsentrydescplural{cgf} (\glsentryplural{cgf})}
}
\newglossaryentry{pdf}
{
  name={PDF},
  description={probability density function},
  first={\glsentrydesc{pdf} (\glsentrytext{pdf})},
   plural={PDF's},
  descriptionplural={probability density functions},
  firstplural={\glsentrydescplural{pdf} (\glsentryplural{pdf})}
}

\newglossaryentry{ccdf}
{
  name={CCDF},
  description={complementary cumulative distribution function},
  first={\glsentrydesc{ccdf} (\glsentrytext{ccdf})}
}

\newglossaryentry{cdf}
{
  name={CDF},
  description={cumulative distribution function},
  first={\glsentrydesc{cdf} (\glsentrytext{cdf})}
}

\newglossaryentry{gmi}
{
  name={GMI},
  description={generalized mutual information},
  first={\glsentrydesc{gmi} (\glsentrytext{gmi})}
}

\newglossaryentry{arq}
{
  name={ARQ},
  description={automatic repeat request},
  first={\glsentrydesc{arq} (\glsentrytext{arq})}
}

\newglossaryentry{harq}
{
  name={HARQ},
  description={hybrid automatic repeat request},
  first={\glsentrydesc{harq} (\glsentrytext{harq})}
}

\newglossaryentry{fbl}
{
  name={FBL-NF},
  description={fixed blocklength with no feedback},
  first={\glsentrydesc{fbl} (\glsentrytext{fbl})}
}
\newglossaryentry{ssnf}
{
  name={SS-NF},
  description={single-shot no-feedback},
  first={\glsentrydesc{ssnf} (\glsentrytext{ssnf})}
}

\newacronym{urllc}{URLLC}{ultra-reliable low-latency communications}

\newglossaryentry{vlsf}
{
  name={VLSF},
  description={Variable-length stop-feedback},
  first={\glsentrydesc{vlsf} (\glsentrytext{vlsf})}
}

\newglossaryentry{vlnsf}
{
  name={VLNSF},
  description={variable-length noisy stop feedback},
  first={\glsentrydesc{vlnsf} (\glsentrytext{vlnsf})}
}

\newglossaryentry{dmt}
{
  name={DMT},
  description={diversity-multiplexing tradeoff},
  first={\glsentrydesc{dmt} (\glsentrytext{dmt})}
}

\newglossaryentry{dmdt}
{
  name={DMDT},
  description={diversity-multiplexing-delay tradeoff},
  first={\glsentrydesc{dmdt} (\glsentrytext{dmdt})}
}

\newglossaryentry{tddt}
{
  name={TDDT},
  description={throughput-diversity-delay tradeoff},
  first={\glsentrydesc{tddt} (\glsentrytext{tddt})}
}

\newglossaryentry{tdd}
{
  name={TDD},
  description={time-division duplex},
  first={\glsentrydesc{tdd} (\glsentrytext{tdd})}
}

\newglossaryentry{stbc}
{
  name={STBC},
  description={space-time block-codes},
  first={\glsentrydesc{stbc} (\glsentrytext{stbc})},
  plural={STBC's},
  descriptionplural={space-time block-codes},
  firstplural={\glsentrydescplural{stbc} (\glsentryplural{stbc})}
}

\newglossaryentry{harqir}
{
  name={HARQ-IR},
  description={hybrid automatic repeat request with incremental redundancy},
  first={\glsentrydesc{harqir} (\glsentrytext{harqir})}
}

\newglossaryentry{harqcc}
{
  name={HARQ-CC},
  description={hybrid automatic repeat request with chase combining},
  first={\glsentrydesc{harqcc} (\glsentrytext{harqcc})}
}

\newglossaryentry{wp1}
{
  name={w.p.1.},
  description={with probability one},
  first={\glsentrydesc{wp1} (\glsentrytext{wp1})}
}

\newglossaryentry{vlff}
{
  name={VLFF},
  description={variable-length full feedback},
  first={\glsentrydesc{vlff} (\glsentrytext{vlff})}
}

\newglossaryentry{dmc}
{
  name={DMC},
  description={discrete memoryless channel},
  first={\glsentrydesc{dmc} (\glsentrytext{dmc})}
  plural={DMCs},
  descriptionplural={discrete memoryless channels},
  firstplural={\glsentrydescplural{dmc} (\glsentryplural{dmc})}
}

\usepackage{pgfplots}
\pgfplotsset{compat=1.14}
\usepgfplotslibrary{groupplots}
\usetikzlibrary{pgfplots.groupplots} 
\usepackage{subcaption}
\let\abs\undefined
\newcommand{\abs}[1]{\lvert#1\rvert}		

\def\tr{\mathrm{tr}}

\def\diag{\mathrm{diag}}

\newtheorem{theorem}{Theorem}

\newtheorem{remark}{Remark}

\def\tr{\mathrm{tr}}
\def\Qexp{\Psi_{n,\zeta}}

\def\diag{\mathrm{diag}}

\def\tr{\mathrm{tr}}

\def\Htran{\mbox{\tiny $\mathrm{H}$}}
\def\Ttran{\mbox{\tiny $\mathrm{T}$}}
\def\bphiu{\boldsymbol{\phi}} 

\begin{document}

\title{Cell-Free Massive MIMO for URLLC:\\ A Finite-Blocklength Analysis}

\author{Alejandro Lancho,~\IEEEmembership{Member,~IEEE}, Giuseppe~Durisi,~\IEEEmembership{Senior Member,~IEEE}, and Luca Sanguinetti,~\IEEEmembership{Senior Member,~IEEE}

\thanks{Alejandro Lancho is with the Department of Electrical Engineering and Computer Science, Massachusetts Institute of Technology, Cambridge 02139, MA, USA (e-mail: lancho@mit.edu). Giuseppe Durisi is with the Department of
Electrical Engineering, Chalmers University of Technology, Gothenburg 41296,
Sweden (e-mail: durisi@chalmers.se). Luca Sanguinetti is with
the Dipartimento di Ingegneria dell'Informazione, University of Pisa, 56122
Pisa, Italy (e-mail: luca.sanguinetti@unipi.it).} 
\thanks{Alejandro Lancho has received funding from the European Union's Horizon 2020 research and innovation programme under the Marie Sklodowska-Curie grant agreement No. 101024432.
This work was partly supported by the Swedish Research Council under grant
2021-04970 and by the Wallenberg AI, autonomous systems, and software program.
L. Sanguinetti was partially supported by the Italian Ministry of Education and Research (MIUR) in the framework of the FoReLab projects (Departments of Excellence) and by the University of Pisa under the ``PRA - Progetti di Ricerca di Ateneo'' (Institutional Research Grants) - Project no. PRA-2022-2023-91 ``INTERCONNECT''. This work is also supported by the National Science Foundation under Grant No CCF-2131115.}
\thanks{The material in this paper was presented in part at the IEEE Int. Workshop Signal
Process. Advances Wireless Commun. (SPAWC), Lucca, Italy, Sep. 2021.~\cite{lancho21-07p}.}
}
\maketitle

\begin{abstract}
    We present a general framework for the characterization of the packet error
    probability achievable in cell-free Massive multiple-input multiple output
    (MIMO) architectures deployed to support ultra-reliable low-latency (URLLC)
    traffic. The framework is general and encompasses both centralized and
    distributed cell-free architectures, arbitrary fading channels and channel
    estimation algorithms at both network and user-equipment (UE) sides, as well
    as arbitrary combining and precoding schemes. 
    The framework is used to
    perform numerical experiments on specific scenarios, which illustrate the superiority of cell-free
    architectures compared to cellular architectures in
    supporting URLLC traffic in uplink and downlink. 
    Also, these numerical experiments 
    provide the following insights into the design of cell-free
    architectures for URLLC: \emph{i}) minimum mean square error (MMSE) spatial processing must be used to
    achieve the URLLC targets; \emph{ii}) for a given total number of antennas per coverage
    area, centralized cell-free solutions involving single-antenna access points
    (APs) offer the
    best performance in the uplink, thereby highlighting the importance of reducing
    the average distance between APs and UEs in the URLLC regime; \emph{iii}) this observation applies also to the downlink, provided that the
    APs transmit
    precoded pilots to allow the UEs to estimate
    accurately the precoded channel. 
\end{abstract}

 \smallskip

\section{Introduction}\label{sec:intro}
One of the most challenging use cases in next-generation
 wireless communication systems (5G and beyond) is \gls{urllc}---a novel use
case aimed at providing connectivity to real-time mission-critical applications~\cite{3GPP22.104}. 
In some of the most challenging \gls{urllc} scenarios, such as factory
automation, small information payloads, on the order of $100$ bits, must be delivered
within hundreds of microseconds and with a reliability no smaller than
$99.999\%$. Under the stringent latency requirements of \gls{urllc} services and
applications, exploiting time diversity is not possible. Furthermore, the use of
frequency diversity is problematic, especially in the \gls{ul}, where current
standardization rules do not allow \glspl{ue} to spread a coded packet over
noncontiguous frequency resources. Hence, exploiting space diversity
becomes crucial to fulfill the high reliability constraints required in
\gls{urllc}. This can be achieved by the use of Massive \gls{mimo}~\cite{Marzetta10}, which is a key technology in 5G, owing to its ability to
substantially increase the spectral
efficiency~\cite{Ngo13,Sanguinetti18,bjornson19-b} and the energy
efficiency~\cite{bjornson15-a} of cellular networks. Massive \gls{mimo} refers
to a wireless network where \glspl{bs}, equipped with a very large number $M$ of
antennas, serve a multitude of \glspl{ue} via linear spatial signal
processing~\cite{bjornson19}. Important challenges in Massive \gls{mimo} are
the large pathloss variations and the inter-cell interference, in particular for the
cell-edge \glspl{ue}~\cite{Sanguinetti20}. These two phenomena may limit the 
overall network performance and prevent the use of Massive MIMO to support URLLC services. 

An alternative network structure,
known as cell-free Massive \gls{mimo}, was recently proposed to overcome these
issues~\cite{ngo17-03a,nayebi17-07x}. In this type of network, all the \glspl{ue}
in a large coverage area are jointly served by multiple distributed \glspl{ap}.
The fronthaul connections between the \glspl{ap} and a \gls{cpu} enable the
division of the processing tasks needed to coherently serve all the active
\glspl{ue}.

The majority of existing literature on 
cell-free Massive \gls{mimo} has been so far focused on the \textit{ergodic
regime}, where the propagation channel evolves according to a block-fading
model, and one focuses on the asymptotic limit in which, as the codeword length
goes to infinity, the codeword spans an arbitrarily large number of fading blocks.
Unfortunately, these assumptions are highly questionable in
\gls{urllc} scenarios~\cite{durisi16-09a}.
Hence, it is unclear  whether the design
guidelines obtained so far for cell-free Massive
MIMO~\cite{bjornson20-1a,Demir21} apply to \gls{urllc} traffic.

A similar issue, but for the case of Massive \gls{mimo} cellular networks, was
recently addressed in~\cite{ostman20-09b}, where the authors presented a general 
framework, built on rigorous finite-blocklength information-theoretic bounds and
approximations~\cite{polyanskiy10-05a,martinez11-02a,scarlett14-05a}, 
to characterize the performance attainable in \gls{urllc} scenarios. 

The main goal of this paper is to extend the framework 
introduced in~\cite{ostman20-09b} to the case of cell-free Massive \gls{mimo}, and to
use this framework to perform numerical experiments that shed light into the
design of cell-free Massive \gls{mimo} architectures able to support \gls{urllc} traffic in
\gls{ul} and \gls{dl}.
\subsection{State of the Art}
Several attempts to account for finite-blocklength effects when characterizing the
performance of cellular and cell-free Massive MIMO networks can be found in the
literature. For example, the authors of~\cite{Karlsson18,bana18-10a} assumed
that the fading channel stays constant during the transmission of a codeword
(the so-called quasi-static fading scenario) and used the outage
capacity as asymptotic performance metric to characterize
the performance of cellular Massive MIMO. 
As illustrated in~\cite{ostman20-09b}, although the
quasi-static fading scenario is relevant for \gls{urllc},  the 
infinite-blocklength assumption implicit in outage analyses may yield significantly 
incorrect estimates of the error probability. 
Another drawback resulting from the use of the outage-capacity framework is
that it is generally not possible to account for the \gls{csi} acquisition
overhead, which is, however, significant in the \gls{urllc}
regime~\cite{ostman19-02}. 
Indeed, quasi-static channels can be learned perfectly at the receiver in the
asymptotic limit of large blocklength by simply transmitting a number of pilot
symbols that grows sublinearly with the blocklength. 
The attempts made so far to
include channel-estimation overhead in the outage
setup~\cite{Karlsson18,bana18-10a} are not convincing  from a theoretical
perspective, since they partly rely on results that hold only in
the ergodic setting.

A theoretically satisfying framework that  
results in practically relevant information-theoretic upper bounds on the error
probability was introduced in~\cite{ostman20-09b}. 
This framework encompasses the use of a mismatch receiver that treats the channel estimate,
obtained using a fixed number of pilot symbols, as perfect, and it relies on the
finite-blocklength tools developed in~\cite{polyanskiy10-05a,martinez11-02a} and later
extended to wireless fading channels in, e.g.,~\cite{yang14-07c,durisi16-02a,ostman19-02}.

Finite-blocklength analyses have been recently conducted for both cellular Massive
\gls{mimo} networks~\cite{Zeng2020, Ren2020,Zeng20-01b} and cell-free Massive \gls{mimo}
networks~\cite{nasir21-04u}. The analysis in these papers relies on the so-called \textit{normal approximation}~\cite[Eq.~(291)]{polyanskiy10-05a}, which,
although capturing finite-blocklength effects, tends to suffer from low accuracy
for the error probabilities that are of interest in \gls{urllc}~\cite{ostman20-09b}.
Furthermore, the use of the normal approximation for the case of imperfect
\gls{csi} in both~\cite{Zeng2020} and~\cite{Ren2020} is not convincing, since the
provided approximation does not depend on the instantaneous channel estimation error, but only
on its variance. This is not compatible with a scenario in which the
channel stays constant over the duration of each codeword.
The analysis conducted in~\cite{nasir21-04u} is limited to the case in which
perfect  \gls{csi} is available at the receiver, which, as already mentioned,
is not a reasonable assumption in the \gls{urllc} regime.
\subsection{Contributions}
We present a general framework for characterizing in a numerical efficient way
the packet error probability
achievable in cell-free Massive MIMO architectures supporting URLLC services.
Our framework, which generalizes to cell-free Massive MIMO architectures the one
presented in~\cite{ostman20-09b}
for cellular Massive MIMO architectures, is applicable to both centralized and
distributed cell-free systems, and allows for arbitrary fading distributions,
channel estimation schemes, and spatial combining/precoding processing. We use the proposed framework to conduct numerical experiments pertaining an
automated-factory deployment scenario, with the objective to compare the
performance of cellular and cell-free Massive MIMO architectures and to draw
guidelines for the design of cell-free architectures providing URLLC services.
As performance metric, we use the network availability, which we define as the
fraction of UE placements for which the per-link error probability, averaged
over the small-scale fading and the additive noise, is below the URLLC target. The following conclusions can be drawn from our numerical experiments:
\begin{enumerate}[(i)]
    \item  \gls{mmse} spatial processing is necessary to achieve the \gls{urllc} reliability
  requirements, independently of the chosen architecture.
\item  For a fixed total number of antennas per coverage area, centralized cell-free
  architectures with single-antenna APs offer the best \gls{ul} performance.
  Indeed, in the URLLC regime it is beneficial to minimize the average distance
  between UEs and APs by densifying the AP deployment. The use of multiple
  antennas at the APs is not beneficial, if it results in a reduction of the
  number of deployed APs.
\item  For this observation to apply to the \gls{dl} as well, precoded \gls{dl} pilots
  need to be transmitted by the APs to allow the UE to acquire channel knowledge.
  Without \gls{dl} pilots, the \gls{dl} performance of the centralized cell-free
  architecture degrades significantly, and becomes inferior to that of a
  distributed cell-free architecture with multiple-antenna APs.
\end{enumerate}
We notice that the above conclusions also hold true in the ergodic
regime when the average sum spectral efficiency of cell-free networks is
considered. In this context, the first and second observations have been
previously reported in, e.g.,\cite{Demir21,bjornson20-1a} while the third
one can be found in, e.g.,~\cite{interdonato19-08}. Our numerical experiments
not only confirm these previous findings but further reinforce them. For
example, while in~\cite{Demir21,bjornson20-1a} the MMSE processing was
shown to be the preferable option for cell-free networks, our analysis shows
that it is \emph{mandatory} to satisfy the stringent requirements of URLLC
scenarios. The same is valid for the use of downlink pilots.
\subsection{Paper Outline}
In Section~\ref{sec:fbl-intro}, we present the proposed finite-blocklength framework, and
we use it  to illustrate, for a simple
nonfading scenario, the benefits of
cell-free architectures over cellular architectures for URLLC services.
In Section~\ref{sec:mimo}, we introduce a general finite-blocklength system model for
cell-free Massive \gls{mimo} networks. This is used in Section~\ref{sec:ulDataPhase} to detail the \gls{ul} and \gls{dl} of different network architectures.  
The preliminary analysis reported
in Section~\ref{sec:fbl-intro} is then extended to the more practically relevant fading networks considered in
Section~\ref{sec:numericalResult}. 
Some conclusions are drawn in Section~\ref{sec:conclusions}.
\subsection{Notation}
Lower-case bold letters are used for vectors and upper-case bold letters are used for matrices.  The
circularly-symmetric Gaussian distribution is denoted by $\jpg(0,\sigma^2)$,
where $\sigma^2$ is its variance. 
We use $\Ex{}{\cdot}$ to indicate the expectation operator, $\mathbb{V}[\cdot]$ to indicate the variance operator, and $\prob{\cdot}$
for the probability of an event. 
The natural logarithm is denoted by $\log(\cdot)$, and $Q(\cdot)$ stands for the Gaussian $Q$-function.  
The operators $(\cdot)^{\Ttran}$, $(\cdot)^*$, and $(\cdot)^{\Htran}$ denote transpose, complex conjugate, and Hermitian transpose,
respectively.  The Euclidean norm is denoted by $\vecnorm{\cdot}$.
\subsection{Reproducible Research} The Matlab code used to obtain the simulation results is available at:
\url{https://github.com/infotheorychalmers/URLLC_cell-free_Massive_MIMO}.


%
\section{Review of a Finite-Blocklength Upper-Bound \\on the Error Probability}\label{sec:fbl-intro}
\subsection{Desirable Features}\label{sec:features}
A finite-blocklength information-theoretic framework to characterize the performance 
achievable in multiuser Massive \gls{mimo} systems, be them
cellular or cell-free, needs to capture the following
aspects: 
\begin{itemize} 
    \item It must allow for linear spatial processing, used to separate
    the signals generated by/intended to the different \glspl{ue}.
    \item It must allow for pilot-based \gls{csi} acquisition and apply to the practically relevant scenario
        in which decoding is performed under the assumption that the acquired \gls{csi} is exact.
    \item It must apply to a scenario in which the additive noise term includes
        not only thermal noise, but also residual
        multiuser interference after spatial processing. 
\end{itemize}

We proceed as in~\cite{ostman20-09b} and start by considering
the simple case in which the received signal is the superposition of a scaled
version of the desired signal and additive Gaussian noise. We present a
finite-blocklength upper bound on the error probability for this simplified
channel model (Section~\ref{sec:FBL_bound}), and describe an efficient method
for its numerical evaluation (Section~\ref{sec:saddle}), based on the
saddlepoint approximation [24, Ch. XVI]. 
The simple channel model introduced in this section constitutes
the building block for the analysis of the error probability achievable in
cell-free Massive MIMO networks. In
Section~\ref{sec:benefit_cell_free}, we use the finite-blocklength bound, suitably adapted to cell-free 
networks, to exemplify the potential gains provided by cell-free architectures over cellular
architectures on a simplified nonfading setup. 
This will motivate the more thorough studies
performed in Section~\ref{sec:numericalResult}. 
We emphasize that the finite-blocklength upper bound presented in
Section~~\ref{sec:FBL_bound} and its numerically efficient saddlepoint approximation presented in Section~~\ref{sec:saddle} coincide with the ones reported in~\cite{ostman20-09b}.
However, their adaptation (see Sections~~\ref{sec:mimo}
and~\ref{sec:ulDataPhase}) to both centralized and distributed cell-free architectures, as well as to the case of precoded DL pilot transmission, is novel. 
Indeed, the analysis in~\cite{ostman20-09b} is limited to cellular Massive MIMO
networks. Furthermore, even though precoded DL pilots have been considered in
the literature (see, e.g., \cite{interdonato19-08}), to the best of our
knowledge, this is the first time they are considered in the context of
short-packet communications.

\subsection{Random-coding union bound}\label{sec:FBL_bound}
A framework satisfying the requirements listed in Section~\ref{sec:features} can be obtained via the so-called
\gls{rcus} introduced in~\cite{martinez11-02a}.
To introduce this bound, let us consider the following scalar input-output relation:
\begin{equation}\label{eq:simplified_channel}
  v[k] = g q[k] + z[k], \quad k=1,\dots,n.
\end{equation}
Here, $q[k]$ denotes the $k$th entry of the length-$n$ codeword transmitted by a given user, $v[k]$ is the
corresponding received signal after linear processing, $g$ denotes the effective channel after linear
processing, which we assume to stay constant over the duration of the codeword, and $z[k]$ is the additive noise
signal, which includes also the residual multiuser interference. Note that, in this paper, we use the packet length $n$ (also referred to as blocklength) as a proxy for the latency experienced in the access part of the wireless network. 
We do not model in our analysis delays due to user scheduling, nor the additional
delays experienced in the transmission/processing of the information data over
the fronthaul connecting the \glspl{ap} to the \gls{cpu}.

To derive the bound, we shall assume that the receiver does not know $g$, but has access to an estimate
$\widehat{g}$ that is treated as perfect. This estimate may be obtained via pilot transmission, or may simply be
based on the knowledge of first-order statistics of $g$. 
The first situation is relevant in the \gls{ul}, whereas the second situation typically occurs in the \gls{dl} (see,
e.g.,~\cite{bjornson19}).

To estimate the transmitted codeword, the decoder performs \gls{snn} decoding, i.e., it seeks the codeword
that, after being scaled by the estimated channel gain $\widehat{g}$, is closest to the received vector.
Mathematically, the decoder solves the following problem:
\begin{equation}\label{eq:mismatched_snn_decoder}
  \widehat \vecq=\argmin_{\widetilde \vecq\in\mathcal{C}} \vecnorm{\vecv-\widehat{g}\widetilde \vecq}^2.
\end{equation}
Here, $\vecv=[v[1],\dots,v[n]]^{\Ttran}$, the vector $\widehat \vecq$ stands for the codeword chosen by the decoder, and $\mathcal{C}$ denotes the set of length-$n$ codewords.

The \gls{rcus} provides a random coding bound on the packet error probability $\epsilon=\prob{\widehat \vecq\neq
\vecq}$  achieved when the decoder operates according to the rule~\eqref{eq:mismatched_snn_decoder}. The
following theorem provides such a bound for the case of Gaussian codebooks,  i.e., codebooks with codeword
entries generated independently from a $\jpg(0,\rho)$ distribution. 

\begin{theorem}[\!\!{\cite[Th.~1]{ostman20-09b}}]\label{thm:rcus}
  Assume that $g\in\mathbb{C}$ and $\widehat{g}\in\mathbb{C}$ in~\eqref{eq:simplified_channel} are random
  variables drawn according to an arbitrary joint distribution. For all $s>0$, there exists a coding scheme with $m$
  codewords of length $n$ operating according to the mismatched \gls{snn} decoding
  rule~\eqref{eq:mismatched_snn_decoder}, whose error probability $\epsilon$ is upper-bounded by
  \begin{IEEEeqnarray}{lCl} 
    \epsilon &=& \prob{\widehat \vecq\neq \vecq}\nonumber\\
     &\leq& \Ex{g,\widehat{g}}{\prob{\sum_{k=1}^n {\imath_s(q[k],v[k])} \leq 
    \log\frac{m-1}{u} \bigg\given g, \widehat{g}}}.\IEEEeqnarraynumspace \label{eq:rcus_fading}
  \end{IEEEeqnarray}
 Here, $u$ is a random variable that is uniformly distributed over the interval $[0,1]$ and
 $\imath_s(q[k],v[k])$ is the so-called \emph{generalized information density}, which for the case of Gaussian
 codebooks, is given by
 \begin{multline}
     \imath_s(q[k],v[k]) = -s \left|{v[k] - \widehat{g} q[k]}\right|^2 \\ + \frac{s\abs{v[k]}^2}{1+s\rho\abs{\widehat{g}}^2} + 
  \log\lro{1+s\rho\abs{\widehat{g}}^2}.
\label{eq:simple_infodens}
 \end{multline}
 Finally, the average in~\eqref{eq:rcus_fading} is taken over the joint distribution of $g$ and $\widehat{g}$. 
\end{theorem}
\begin{IEEEproof}
See~\cite[App. A]{ostman20-09b}.
\end{IEEEproof}

\smallskip
\begin{remark}
Note that the bound is valid for all values of $s>0$ and can be tightened by
performing an optimization over this parameter. 
We also highlight that 
bound \eqref{eq:rcus_fading} holds for any channel law $g$ and channel
estimate $\widehat{g}$. The implications of this property will become clear
in Section~\ref{sec:ulDataPhase}. 
In~\eqref{eq:rcus_fading}, we
used the law of total probability to express the RCUs as an average, over
the channel $g$ and its estimate $\widehat{g}$ of a tail probability. This
will turn out convenient in view of the application of the
saddlepoint method, discussed in the next section.
Note that, although the tail probability is conditioned with respect to $g$,
this does not mean that $g$ is revealed to the receiver. 
Indeed, the operation performed by the receiver is fully specified by~\eqref{eq:mismatched_snn_decoder},
which requires only knowledge of $\widehat{g}$.
\end{remark}
\begin{remark}
  Within the context of point-to-point single-antenna transmission, the penalty incurred by the choice of using pilot-aided transmission
  instead of more sophisticated noncoherent schemes that do not require channel estimates  was characterized in~\cite{ostman19-02} using information-theoretic bounds similar to the ones used in this paper. As shown in~\cite{ostman19-02}, one attractive feature of the SNN decoder is that information-theoretic bounds on
  its error probability can be approached in practice using good channel codes for the nonfading AWGN channel.
  In contrast, for the optimal noncoherent ML decoder considered in, e.g.,~\cite{lancho20_04}, approaching information-theoretic error-probability bounds with low-complexity coding schemes is still an open problem (note, however, the recent progress reported in~\cite{yuan21-06l}).
\end{remark}  
\subsection{Saddlepoint approximation}\label{sec:saddle}

Unfortunately, the bound~\eqref{eq:rcus_fading} is difficult to evaluate
numerically. Indeed, the conditional probability inside the expectation
in~\eqref{eq:rcus_fading}  is not known in closed form, and evaluating it
accurately for the error-probabilities of interest in \gls{urllc} is time
consuming. One common approach to simplify its evaluation is to invoke the
Berry-Esseen central limit theorem~\cite[Ch. XVI.5]{feller71-a} and replace the
probability in~\eqref{eq:rcus_fading} with a closed-form approximation that
involves the Gaussian $Q(\cdot)$ function and the first two moments of the
generalized information density, which can be expressed in closed
form.\footnote{Although to apply the Berry-Esseen theorem one needs to
verify that the third absolute central moment is bounded, to evaluate the normal
approximation one needs only to compute the first two moments, since the term
involving the third absolute central moment is hidden in the $\landauO(1
\sqrt{n})$ term in~\eqref{eq:normal_approximation}.} The resulting
approximation is given as
\begin{multline}
  \prob{\sum_{k=1}^n {\imath_s(q[k],v[k])} \leq \log\frac{m-1}{u}}\\ =
Q\lefto(\frac{nI_s-\log(m-1)}{\sqrt{nV_s}}\right) + \landauO\lefto(\frac{1}{\sqrt{n}}\right)\label{eq:normal_approximation}
\end{multline}
where $I_s=\Ex{}{\imath_s(q[1],v[1])}$ is the so-called \emph{generalized mutual
information}~\cite[Sec.~III]{lapidoth02-05a},
\begin{equation}
  V_s=\Ex{}{\abs{\imath_s(q[1],v[1])-I_s}^2}
\end{equation}
is the variance of the information density, typically referred to as \emph{channel
dispersion}~\cite[Sec.~IV]{polyanskiy10-05a}, and $\landauO\lefto({1}/{\sqrt{n}}\right)$ accounts for terms
that decay no slower than $1/\sqrt{n}$ as $n\to\infty$.
Neglecting the $\landauO\lefto({1}/{\sqrt{n}}\right)$ term in~\eqref{eq:normal_approximation},
we obtain the so-called \emph{normal approximation}. 
As shown in~\cite[Fig.~1]{ostman20-09b}, this approximation is accurate only when the rate $R=(\log m)/n$ is close to
$I_{s}$~\cite{lancho20_04}, as a consequence of the central limit
theorem~\cite[Ch.~5.11]{grimmett01}. 
Unfortunately, this is typically not the case
for the low error-probabilities of interest in \gls{urllc}.

An alternative approximation, which turns out to be accurate for a much larger range of error-probability
values, including the ones of interest in \gls{urllc}, can be obtained using the so-called \emph{saddlepoint
method}~\cite[Ch. XVI]{feller71-a}. The resulting approximation for the conditional probability
inside the expectation in~\eqref{eq:rcus_fading} is also in closed form for the setup
considered in the present paper. 
As a consequence, the resulting approximation, commonly referred to as saddlepoint approximation, has essentially the same
computational complexity as the normal approximation. A detailed
    analysis of the complexity and the accuracy of various types of normal and
    saddlepoint approximations was recently reported 
in~\cite{kislal22-11a}. 
For the setup considered in this section, the saddlepoint
approximation was previously derived in \cite[Th. 2]{ostman20-09b}. 
We provide it in Theorem~\ref{thm:saddlepoint} below for completeness and emphasize that this is not a
novel contribution of this paper.
\begin{theorem}\label{thm:saddlepoint}
  Let $m=e^{nR}$ for some $R>0$, and define $\kappa(\zeta)$, $\kappa''(\zeta)$ and $\kappa''(\zeta)$ as
   \begin{IEEEeqnarray}{lCl}
  \kappa(\zeta) 
     &=&{}
    -\zeta\log\lro{1+s\rho\abs{\widehat{g}}^2}\nonumber\\
    &&\qquad{} 
    - \log \lro{1+\lro{\beta_B-\beta_A}\zeta -\beta_A\beta_B(1-\nu)\zeta^2}\IEEEeqnarraynumspace\label{eq:cgf}\\
    \kappa'(\zeta) &=&{}
    -\log\lro{1+s\rho\abs{\widehat{g}}^2} \nonumber\\
    &&\qquad{} 
    - \frac{\lro{\beta_B-\beta_A} -2\beta_A\beta_B(1-\nu)\zeta}{1+\lro{\beta_B-\beta_A}\zeta -\beta_A\beta_B(1-\nu)\zeta^2} \label{eq:cgf_1} \\
    \kappa''(\zeta) &=& \lrho{\frac{\lro{\beta_B-\beta_A} -2\beta_A\beta_B(1-\nu)\zeta}{1+\lro{\beta_B-\beta_A}\zeta -\beta_A\beta_B(1-\nu)\zeta^2}}^2 \nonumber\\
    &&\qquad{} 
    +  \frac{2\beta_A\beta_B(1-\nu)}{1+\lro{\beta_B-\beta_A}\zeta -\beta_A\beta_B(1-\nu)\zeta^2} \label{eq:cgf_2}
\end{IEEEeqnarray} 
   where 
  \begin{IEEEeqnarray}{lCl}
    \beta_A &=& s(\rho \abs{g-\widehat{g}}^2+\sigma^2)\label{eq:betaA}\\
    \beta_B &=& \frac{s}{1+s\rho\abs{\widehat{g}}^2} \lro{\rho\abs{g}^2 + \sigma^2}\label{eq:betaB}\\
    \nu 
    &=&\frac{s^2 \left|{\rho  \abs{g}^2 + \sigma^2-g^*\widehat{g}\snr}\right|^2}{\beta_A \beta_B (1+s \rho \abs{\widehat{g}}^2)}.
    \label{eq:corr_coeff_SISO}
  \end{IEEEeqnarray}
  One can show that $I_{s}=-\kappa'(0)$.   
  Let $R_s^{\mathrm{cr}}=-\kappa'(1)$  and let $\zeta\in(\underline{\zeta},\overline{\zeta})$ be the solution to the
  equation $R=-\kappa'(\zeta)$.  If $\zeta\in[0,1]$, then 
  \begin{multline}
     \prob{\sum_{k=1}^n {\imath_s(q[k],v[k])} \leq \log\frac{e^{nR}-1}{u}} \\ = e^{n[\kappa(\zeta)+\zeta R]}\lrho{\Qexp\lro{\zeta}+\Qexp\lro{1-\zeta}+o\lro{\frac{1}{\sqrt{n}}}} \label{eq:saddlepoint_U_pos}
  \end{multline}
  where 
  \begin{IEEEeqnarray}{rCl}
    \Qexp\lro{u} & \triangleq & e^{n\frac{u^2}{2}\kappa''(\zeta)}Q\lro{u\sqrt{n\kappa''(\zeta)}}\IEEEeqnarraynumspace\label{eq:help_fcn_theta}
  \end{IEEEeqnarray}
  and $o(1/\sqrt{n})$ comprises terms that vanish faster than $1/\sqrt{n}$ and are uniform in $\zeta$.
  If $\zeta>1$, then 
  \begin{multline}
    \prob{\sum_{k=1}^n {\imath_s(q[k],v[k])} \leq \log\frac{e^{nR}-1}{u}} \\= e^{n[\kappa(1)+ R]}\lrho{\widetilde{\Psi}_n(1,1)+\widetilde{\Psi}_n(0,-1)+\mathcal{O}\lro{\frac{1}{\sqrt{n}}}} \label{eq:saddlepoint_U_pos_cr}
  \end{multline}
  where 
  \begin{multline}
    \widetilde{\Psi}_n(a_1,a_2)\\ = e^{na_1\lrho{R_s^{\mathrm{cr}}-R+\frac{\kappa''(1)}{2}}} Q\lro{a_1\sqrt{n\kappa''(1)}+a_2\frac{n(R_s^{\mathrm{cr}}-R)}{\sqrt{n\kappa''(1)}}}\label{eq:help_fcn_theta_2}
  \end{multline}
  and $\mathcal{O}(1/\sqrt{n})$ comprises terms that decay no slower than $1/\sqrt{n}$ and are uniform in $\zeta$. 
  If $\zeta<0$, then 
  \begin{IEEEeqnarray}{lCl}
    \IEEEeqnarraymulticol{3}{l}{
    \prob{\sum_{k=1}^n {\imath_s(q[k],v[k])} \leq \log\frac{e^{nR}-1}{u}}}\nonumber\\ \quad
    &=& 1 - e^{n[\kappa(\zeta)+\zeta R]}\biggl[ \Qexp(-\zeta) - \Qexp\lro{1-\zeta}  
     +o\lro{\frac{1}{\sqrt{n}}}\biggr].\nonumber\\\label{eq:saddlepoint_U_neg}
  \end{IEEEeqnarray}
\end{theorem}
\subsection{The benefit of cell-free networks over cellular
networks in the finite-blocklength regime}\label{sec:benefit_cell_free} We now use the
bound in~\eqref{eq:rcus_fading} together with the saddlepoint approximation from
Theorem~\ref{thm:saddlepoint} to exemplify the benefits of cell-free networks
over conventional cellular networks, in the \gls{urllc} regime.
Following~\cite[Sec. 1.3]{Demir21}, we compare three setups. The first one is a
single cell with a $64$-antenna Massive \gls{mimo} \gls{ap}; the second one
consists of $64$ small cells served each by a single-antenna \gls{ap}, deployed
on a square grid; the last one is a cell-free network where the same $64$
\gls{ap} locations are used. In the small-cell network, each \gls{ue} is associated to the \gls{ap} providing the best performance. 
We focus on the \gls{ul} and assume that $K$ UEs
are active in the coverage area. We denote by
${\bf h}_k = [h_{k1}, \ldots,h_{kM}]^{\Ttran} \in \mathbb{C}^M$ the channel between UE $k$ and
the $M=64$ antennas or APs. 
The channel gain $\beta(d_{kl})$ (i.e., pathloss or large-scale
fading coefficient) for a propagation distance $d_{kl}$ is modelled
as~\cite[Sec. 1.3]{Demir21}
\begin{equation}\label{eq:large-scale-model-simple}
\beta(d) \,  [\textrm{dB}] = -30.5 - 36.7 \log_{10}\lefto( \frac{d}{1\,\textrm{m}} \right).
\end{equation}
All channels are deterministic (i.e., no fading) and perfectly known to the APs. Detection is performed by using
\gls{mmse} and \gls{mr} combiners in all the three
setups.  Since no fading is present, no averaging over $g$ and $\widehat{g}$
in~\eqref{eq:rcus_fading} is required and the tail probability can be
approximated efficiently using the saddlepoint approximations provided in
Theorem~\ref{thm:saddlepoint}.

We consider a total coverage area of $150$\,m\,$\times\,$\,$150$\,m and drop \glspl{ue} uniformly at random in
the area. We assume that the APs are deployed $10$\m above the UEs. The noise
power is $-96$\,\dBm, which is a reasonable value when the bandwidth is $20$\, MHz.
We consider a codeword length $n=100$ and a rate of $R = 60/100 =0.6$ bits per channel use. 

\begin{figure}[t!]     
  \begin{subfigure}{\columnwidth}
    \centering
    \begin{overpic}[width=0.95\columnwidth]{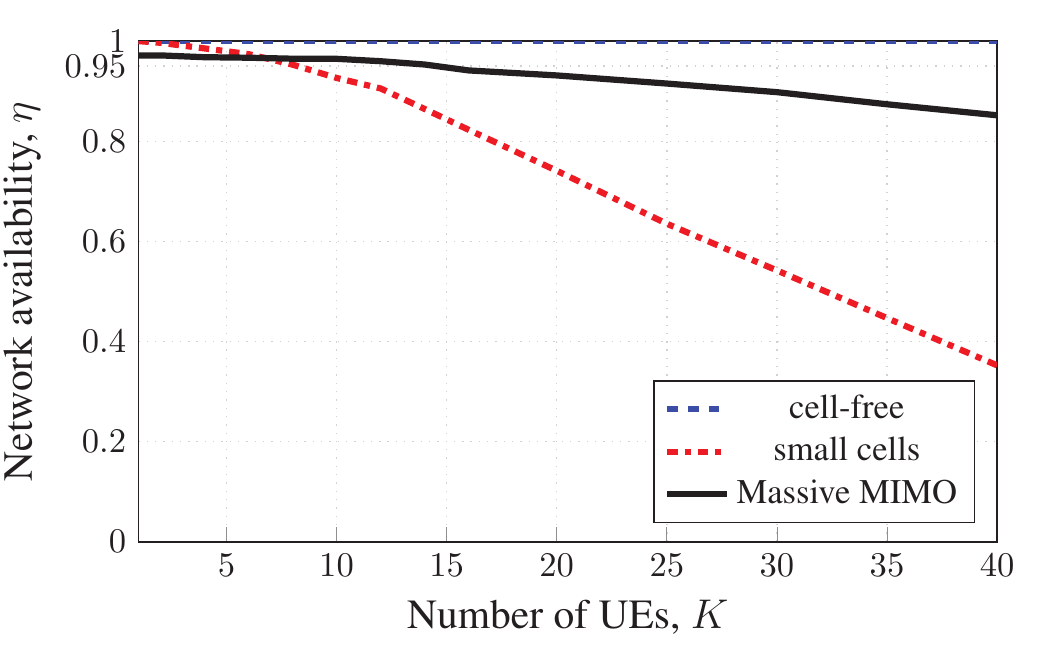}
        		\put(82,53){\footnotesize{$\eta = 0.95$}}
		\put(13,54.5){$- - - - - - - - - - - - - - - - - - - -$}
		\end{overpic}  
   \caption{With \gls{mmse} combiner}
    \label{fig:K_vs_netavail_UL_MMSE_LM_64}
  \end{subfigure}
  
  \begin{subfigure}{\columnwidth}
    \centering
    \begin{overpic}[width=0.95\columnwidth]{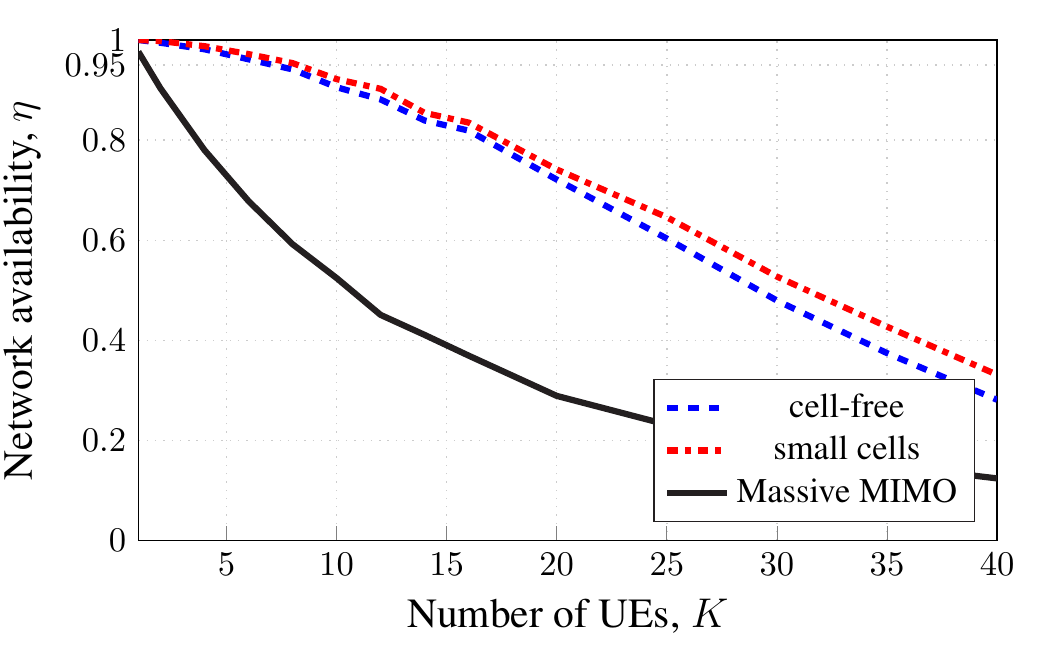}
    		\put(55,53){\footnotesize{$\eta = 0.95$}}
		\put(13,54.5){$- - - - - - - - - - - - - - - - - - - -$}
		\end{overpic}  
   \caption{With \gls{mr} combiner}
    \label{fig:K_vs_netavail_UL_MR_LM_64}
  \end{subfigure}
  \caption{Network availability for each of the three network configurations as a function of the number of active UEs. The total number of antennas or APs is $64$ in all network configurations and the transmit power of each UE is $\rho = -15$\,\dBm. }
  \label{fig:section1_figure1}
\end{figure}

The three different network setups are compared in terms of \emph{network availability}, $\eta$. Following~\cite{ostman20-09b}, this is
defined as the probability, computed with respect to the random \glspl{ue}' positions, that the error
probability is below a given target $\epsilon\sub{target}$, i.e.,
\begin{IEEEeqnarray}{rCl}\label{eq:network_avail}
 \eta = \prob{\epsilon \leq \epsilon\sub{target}}.
\end{IEEEeqnarray}
Fig.~\ref{fig:section1_figure1} shows $\eta$ for
$\epsilon\sub{target} = 10^{-5}$ as a function of 
the number of UEs $K\in\{1,\dots,40\}$.
The transmit power of each \gls{ue} is $\rho = -10$\,\dBm.
We see that, with \gls{mmse} combiner
(Fig.~\ref{fig:K_vs_netavail_UL_MMSE_LM_64}), the cell-free network yields
$\eta =1$ irrespective of $K$.  This is not the case with small cells and Massive
MIMO.  Specifically, the small-cell network yields $\eta = 1$ only when $K=1$.
When $K=20$, the network availability decreases to approximately $0.75$, and it
further decreases to $0.35$ when $K=40$.
Assume now that we are interested in achieving a network availability of $0.95$
(indicated in the figures by a dashed line). This value of $\eta$ is achievable
with small cells only when $K\le 7$.  The cell-free network performs better than
small cells because of its superior ability in managing interference.
Note that Massive MIMO cannot achieve
$\eta=1$ even when $K=1$.
This is due to the larger maximum distance between a randomly placed UE and the BS.
Furthermore, the performance of Massive MIMO decreases, although only
marginally, compared to the small-cell
case, as $K$ increases. With the \gls{mr} combiner (Fig.~\ref{fig:K_vs_netavail_UL_MR_LM_64}), a network
availability $\eta\geq 0.95$ is reached by the cell-free network and by the small-cell network when $K\leq7$, and cannot be achieved by the Massive MIMO
network even when $K=1$. Perhaps surprisingly, the small-cell network slightly
outperforms the cell-free network, which implies that cooperation is only
beneficial when the interference is properly managed via the use of an \gls{mmse}
combiner.
  
\begin{figure}[t!]     
  \begin{subfigure}{\columnwidth}
    \centering
    \begin{overpic}[width=0.95\columnwidth]{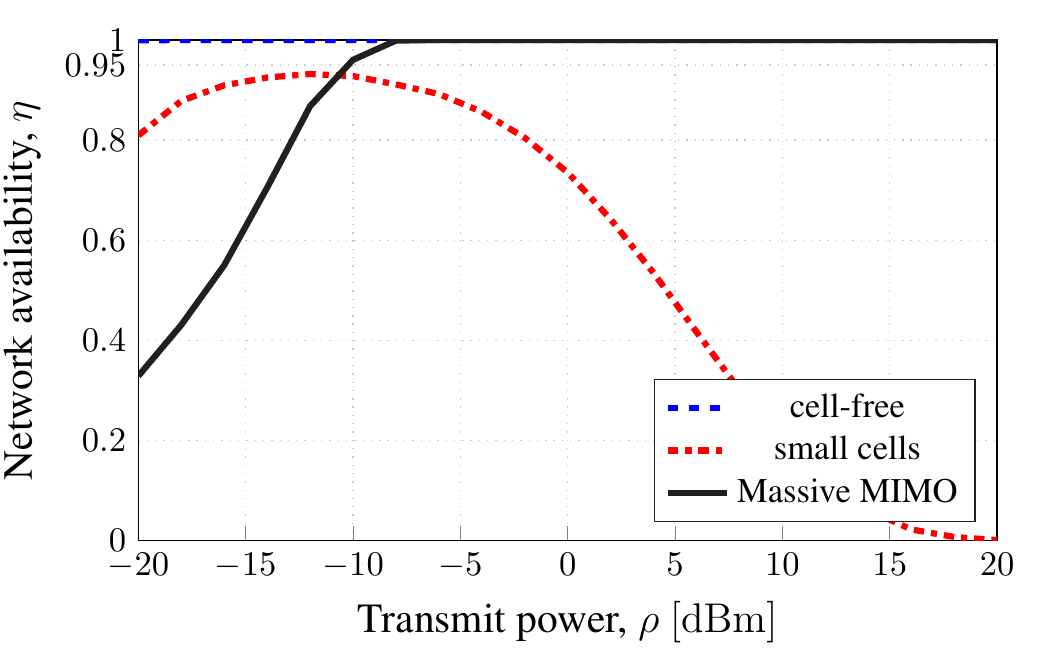}
        		\put(55,53){\footnotesize{$\eta = 0.95$}}
		\put(13,55.4){$- - - - - - - - - - - - - - - - - - - -$}
		\end{overpic}  
   \caption{With \gls{mmse} combiner}
    \label{fig:rho_vs_netavail_UL_MMSE_LM_64}
  \end{subfigure}
  
  \begin{subfigure}{\columnwidth}
    \centering
    \begin{overpic}[width=0.95\columnwidth]{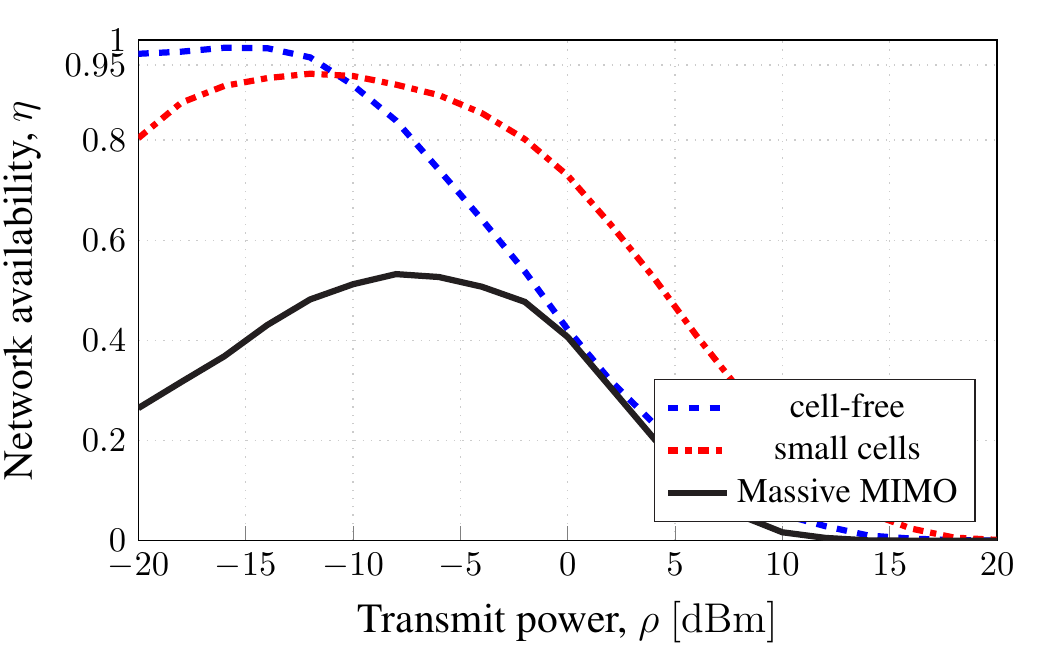}
    		\put(55,53){\footnotesize{$\eta = 0.95$}}
		\put(13,55.4){$- - - - - - - - - - - - - - - - - - - -$}
		\end{overpic}  
   \caption{With \gls{mr} combiner}
    \label{fig:rho_vs_netavail_UL_MR_LM_64}
  \end{subfigure}
  \caption{Network availability for  each of the three network configurations as a
  function of the transmit power of each UE. The total number of antennas or APs is $64$ and the number
  of UEs is $K =10$.}
  \label{fig:section1_figure2}
\end{figure}
In Fig.~\ref{fig:section1_figure2}, we plot $\eta$ as a function of the transmit
power $\rho\in \{-20,\dots, 20\} \dBm$,
$\epsilon\sub{target} = 10^{-5}$ and $K=10$. With
the \gls{mmse} combiner (Fig.~\ref{fig:rho_vs_netavail_UL_MMSE_LM_64}), 
cell-free yields $\eta$ close to $1$ irrespective of $\rho$. On the contrary,
this is possible with
Massive MIMO only when $\rho \ge -8$\,\dBm. A small-cell network performs better
than Massive MIMO for $\rho \le -11$\,\dBm, but its network availability vanishes as
$\rho$ increases and does not achieve $\eta\geq 0.95$. 
Indeed, as the transmit power increases, so does the intercell
interference, which rapidly prevents the network from achieving the target error probability.  With the \gls{mr} combiner (Fig.~\ref{fig:rho_vs_netavail_UL_MR_LM_64}), a
network availability above
$0.95$ can be only achieved by the cell-free network when $\rho\leq -11$\,\dBm, and it cannot be achieved with Massive MIMO or small cells.
Again, the small-cell network outperforms the cell-free
network as the transmit power grows.
This confirms once more  the importance of effective interference management via
the use of \gls{mmse} spatial processing in order to benefit from \gls{ap}
cooperation.

To summarize, from the above analysis, we can conclude that the cell-free
architecture is vastly superior to Massive MIMO and small cells in providing
high network availability for the packet error probabilities of interest in
\gls{urllc}.  However, some strong assumptions were made in the analysis:
nonfading channels, perfect channel state information, and full cooperation
among the APs.  The question thus is: \emph{Can high network availability be
achieved in the UL and DL of practical (centralized or decentralized) cell-free networks where these
assumptions are typically not met?}


\section{Cell-Free Network Model}\label{sec:mimo}
To answer the above question, we now present a more refined system model,
which will allow us to generalize the
observations reported in Section~\ref{sec:benefit_cell_free} to more practically relevant scenarios.
Specifically, we consider a cell-free Massive \gls{mimo} network with $L$ \glspl{ap}, each equipped with $M$
antennas, which are geographically distributed over the coverage area. 
The \glspl{ap} serve jointly $K$
single-antenna \glspl{ue}, and are connected via fronthaul links to the \gls{cpu}.  The standard
time-division duplexing protocol of cellular Massive \gls{mimo} is used, where the $n$ available channel uses
are divided as follows:  $\np^{\rm ul}$ symbols for \gls{ul} pilots;  $n^{\rm ul}$ symbols
for \gls{ul} data;  $\np^{\rm dl}$ symbols for \gls{dl} pilots; and  $n^{\rm dl}$ symbols
for \gls{dl} data.
Note that we explicitly allow in our system model for the use of precoded downlink pilots.
In \gls{urllc} scenarios, their presence will turn out critical  to achieve high network availability in the \gls{dl}.

The channel between \gls{ap} $l$ and \gls{ue} $i$ is denoted by ${\bf h}_{il} \in \mathbb{C}^{M}$. We use a
correlated Rayleigh fading model where ${\bf h}_{il}\sim \jpg({\bf 0}_M,{\bf R}_{il})$ remains constant for
the duration of the $n$ channel uses. The normalized trace $\beta_{il} = \tr({\bf{R}}_{il})/M$ determines
the average large-scale fading between \gls{ap} $l$ and \gls{ue} $i$, while the eigenstructure of
${\bf{R}}_{il}$ describes its spatial channel correlation~\cite[Sec. 2.2]{bjornson19}.
We assume that the channel vectors of different APs are independently distributed;
thus $\mathbb{E}\{\vech_{il}\vech_{il'}^{\rm H}\} = \mathbf{0}_{ML}$, for $l \neq l'$. This is a reasonable
assumption because the APs are typically separated by tens of wavelengths or more. 
The collective channel vector ${\bf
h}_{i} = [{\bf h}_{i1}^{\Ttran} \ldots {\bf h}_{iL}^{\Ttran}]^{\Ttran} \in \mathbb{C}^{ML}$ follows a
$\jpg({\bf 0}_{ML}, {\bf{R}}_{i})$ distribution, where ${\bf{R}}_{i} =
\diag({\bf{R}}_{i1},\ldots,{\bf{R}}_{iL})$. We assume that the spatial correlation matrices
${\bf R}_{il}$ are known to the CPU in a centralized implementation, while each
\gls{ap} has access to its local spatial correlation matrices in a distributed implementation. 
See, e.g., \cite[Sec.~IV]{Sanguinetti20} for practical methods
for estimating spatial correlation matrices.
\subsection{Uplink and Downlink Data Transmission}\label{sec:uldlDataPhase}
In the \gls{ul}, we denote by $x_i^{\rm ul}[k]$ the signal transmitted by \gls{ue} $i$ over channel use $k$. 
The received complex baseband signal ${\bf r}_{l}^{\rm{ul}}[k] \in \mathbb{C}^M$ at \gls{ap} $l$ and channel use
$k\in \{1,\dots,n^{\rm{ul}}\}$ is given by
\begin{equation}\label{eq:ULreceived}
  {\bf r}_{l}^{\rm{ul}}[k] = \sum\limits_{i=1}^{K} \vech_{il} x_i^{\rm ul}[k] + \vecz_{l}^{\rm ul}[k]
\end{equation}
where $\vecz_{l}^{\rm ul}[k]$ denotes the noise vector at AP $l$, which contains i.i.d. entries, distributed
according to $\mathcal{CN}(0,\sigma_{\rm ul}^2)$. 

Similarly, in the \gls{dl} we denote by $x_{i}^{\rm dl}[k]$ the signal intended for \gls{ue} $i$ over channel
use $k \in \{1,\ldots,n^{\rm dl}\}$. 
Let ${\bf w}_{il} \in \mathbb{C}^M$ denote the precoder that \gls{ap} $l$ assigns to \gls{ue} $i$. In the \gls{dl}, the
received signal at \gls{ue} $i$ over channel use $k$ is
\begin{align}
y_i^{\rm{dl}}[k] &= \sum_{l=1}^{L} {\bf h}_{il}^{\Htran} \sum_{j=1}^{K}  {\bf w}_{j l} x_{j}^{\rm dl}[k] + z_i^{\rm dl}[k] \\&
=  {\bf h}_i^{\Htran}{\bf w}_{i} x_{i}^{\rm dl}[k] + {\bf h}_i^{\Htran}\!\!\!\sum_{j=1, j\ne i}^{K}\!\!\!  {\bf w}_{j} x_{j}^{\rm dl}[k] + z_i^{\rm dl}[k] \label{eq:Cell-free-DL}
\end{align}
where ${\bf w}_i = [{\bf w}_{i1}^{\Ttran} \, \ldots \, {\bf w}_{iL}^{\Ttran}]^{\Ttran} \in \mathbb{C}^{ML}$ is
the collective precoding vector, and $z_i^{\rm dl}[k] \sim \jpg(0,\sigma_{\rm dl}^2)$ is the noise at UE $i$. Note also that the precoded channel ${\bf h}_i^{\Htran}{\bf w}_i$ is not Rayleigh distributed. As we remarked in Section \ref{sec:FBL_bound}, this is not a problem since the bound \eqref{eq:rcus_fading} holds for any channel law $g$ and channel estimate $\widehat{g}$.

\subsection{Uplink Pilot Transmission and Channel Estimation}\label{sec:pilots}
The \gls{ul} pilot signature of \gls{ue}~$i$ is denoted by the vector $\bphiu_{i} \in \mathbb{C}^{\np^{\rm
ul}}$ satisfying $\| \bphiu_{i} \|^2  = \np^{\rm ul}$. The elements of $\bphiu_{i}$ are scaled by the
square-root of the pilot power $\sqrt{\rho^{\mathrm{ul}}}$ and transmitted over $\np^{\rm ul}$ channel uses.
At AP $l$, the received pilot signal $ {\bf Y}_l^{\mathrm{pilot}} \in \mathbb{C}^{M \times \np^{\rm ul}}$ is
\begin{IEEEeqnarray}{lCl}
  {\bf Y}_l^{\mathrm{pilot}} = \sqrt{\rho^{\mathrm{ul}}}\sum_{i=1}^K\vech_{il} \bphiu_i^{\Htran} +   {\bf Z}_l^{\mathrm{pilot}}  \label{eq:simo_channel_ul_pilot}
\end{IEEEeqnarray}
where ${\bf Z}_l^{\mathrm{pilot}}\in \mathbb{C}^{M \times \np^{\rm ul}}$ is
noise with independent $\jpg(0,\sigma_{\rm{ul}}^{2})$-distributed elements. 
Since the \gls{cpu} has access to the covariance matrices $\{\mathbf{R}_{il}\}$, 
it can compute the \gls{mmse} estimate of $\vech_{il}$ as \cite[Sec.~3.2]{bjornson19}
 \begin{align} \label{eq:MMSEestimator_h_1}
\widehat \vech_{il}  = \sqrt{\rho^{\mathrm{ul}}}{\bf R}_{il} {\bf Q}_{il} ^{-1}  \left({\bf Y}_l^{\mathrm{pilot}}\bphiu_i\right)
 \end{align}
with $ {\bf Q}_{il}  = \rho^{\mathrm{ul}} \sum_{i'=1}^K{\bf R}_{il} \bphiu_{i'}^{\Htran}\bphiu_i +  \sigma_{\mathrm{ul}}^2  {\bf I}_{M}$. 
We let $\widehat
{\bf h}_{i} = [\widehat {\bf h}_{i1}^{\Ttran} \ldots \widehat {\bf h}_{iL}^{\Ttran}]^{\Ttran}$.
The estimation
error is $\widetilde {\bf h}_{i} = {\bf h}_{i} - \widehat {\bf h}_{i}\sim \jpg ({\bf 0}, {\bf C}_{i} )$ with
${\bf C}_{i} = {\bf R}_{i} - {\bf \Phi}_{i}$ and ${\bf \Phi}_{i} =
\diag({\bf{\Phi}}_{i1},\ldots,{\bf{\Phi}}_{iL})$, where ${\bf{\Phi}}_{il} = \rho^{\rm ul}\np{\bf R}_{il} {\bf Q}_{il}^{-1} {\bf R}_{il}$.

\subsection{Downlink Pilot Transmission and Channel Estimation}\label{sec:DLpilots}
As in the \gls{ul}, we denote the \gls{dl} pilot signature assigned to \gls{ue}~$i$ by the vector $\bphiu_{i}
\in \mathbb{C}^{\np^{\rm dl}}$ satisfying $\| \bphiu_{i} \|^2  = \np^{\rm dl}$. The elements of
$\bphiu_{i}$ are scaled by the square-root of the DL pilot power $\sqrt{\rho^{\mathrm{dl}}}$ and transmitted
over $\np^{\rm dl}$ channel uses. It follows from~\eqref{eq:Cell-free-DL}, that the received signal
${\vecy}_i^{\mathrm{pilot}} \in \mathbb{C}^{1 \times \np^{\rm dl}} $ at UE $i$ is
\begin{align}\notag
{\vecy}_i^{\mathrm{pilot}} 
&= \sqrt{\rho^{\mathrm{dl}}} {\bf h}_{i}^{\Htran} \sum_{j=1}^{K} {\bf w}_{j} \bphiu_{j}^{\Htran} + \vecz_i^{\rm pilot} \\ 
&= \sqrt{\rho^{\mathrm{dl}}} \xi_{ii} \bphiu_{i}^{\Htran} + \sqrt{\rho^{\mathrm{dl}}}  \sum_{j=1, i\ne j}^{K} \xi_{ij} \bphiu_{j}^{\Htran} + \vecz_i^{\rm pilot} \label{eq:simo_channel_dl_pilot}
\end{align}
where $\xi_{ij} =  {\bf h}_{i}^{\Htran} {\bf w}_{j} $ denotes the effective precoded channel to UE $i$ and
${\vecz}_i^{\mathrm{pilot}}\in \mathbb{C}^{1 \times \np^{\rm dl}} $ is noise
with independent 
$\jpg(0,\sigma_{\rm{dl}}^{2})$-distributed
elements. 
The UE multiplies the received row vector ${\vecy}_i^{\mathrm{pilot}}$ with its pilot signature to obtain 
\begin{align}\label{eq:MMSEestimator_observation}
\tilde y_i  = {\vecy}_i^{\mathrm{pilot}} \bphiu_i = \sqrt{\rho^{\mathrm{dl}}} \np^{\rm dl} \xi_{ii}  + \tilde z_i 
\end{align}
with 
\begin{align}
\tilde z_i = \sqrt{\rho^{\mathrm{dl}}}  \sum_{j=1, i\ne j}^{K} \xi_{ij} \bphiu_{j}^{\Htran} \bphiu_i + \vecz_i^{\rm pilot} \bphiu_i.
\end{align}
Since $ \xi_{ii}$ is in general not Gaussian distributed, the \gls{mmse} channel
estimator cannot be expressed in closed form. 
If both the mean and the variance of  $ \xi_{ii}$ are known, one
can utilize the LMMSE estimator, which is given by~\cite[App.~B.4]{bjornson19} 
  \begin{IEEEeqnarray}{lCl}
  \widehat \xi_{ii}  &=& \Ex{}{\xi_{ii}} + \frac{\sqrt{\rho^{\mathrm{dl}}} \np^{\rm dl}\Var{\xi_{ii}}}{\rho^{\mathrm{dl}} (\np^{\rm dl})^2\Var{\xi_{ii}}+ \Var{\tilde z_i }}\nonumber\\
  &&\qquad\qquad\qquad\times \Big(\tilde y_i - \sqrt{\rho^{\mathrm{dl}}} \np^{\rm dl} \Ex{}{\xi_{ii}} - \Ex{}{\tilde z_i }\Big).\IEEEeqnarraynumspace \label{eq:MMSEestimator_h_1_pilots_new}
 \end{IEEEeqnarray}
 An alternative approach is to use the \gls{ls} estimator, which yields 
   \begin{equation} \label{eq:ls_channel_est}
\widehat \xi_{ii}  =  \frac{1}{\sqrt{\rho^{\mathrm{dl}}} \np^{\rm dl} }\tilde y_i.\end{equation}
Unlike the LMMSE estimator, the LS estimator does not require the knowledge of the statistics (i.e., mean and variance) of the precoded channel $\xi_{ii}$, and thus is easier to implement. A similar precoded DL pilot scheme was considered
in~\cite{interdonato19-08} in the context of spectral efficiency analyses of cell-free networks. 
Differently from our analysis, however, the one performed in~\cite{interdonato19-08} considers only distributed cell-free networks, MR precoding, and LMMSE channel estimation.
 
%
%
%

\section{Uplink and Downlink Operation}\label{sec:ulDataPhase}
We now detail the UL and DL of two different implementations of cell-free Massive MIMO, namely, centralized and distributed. 
\subsection{Uplink}
\subsubsection{Centralized Operation}
In the \gls{ul} of a fully centralized implementation, each \gls{ap} $l$ acts only as a remote-radio head,
i.e., as a relay that forwards its received baseband signal ${\bf r}_{l}^{\rm{ul}}[k]$ to the \gls{cpu},
which performs channel estimation and data detection after linear processing. Specifically, to decode the
signal from \gls{ue} $i$, 
the \gls{cpu} computes for $k=1,\ldots,n^{\rm ul}$
\begin{align}
y_i^{\rm{ul}}[k] 
= {\bf u}_{i}^{\Htran} {\bf r}^{\rm{ul}}[k]\label{eq:uplink-CPU-data-estimate}
\end{align} 
where ${\bf u}_{i}= [{\bf u}_{i1}^{\Ttran} \, \ldots \, {\bf u}_{iL}^{\Ttran}]^{\Ttran} \in \mathbb{C}^{ML}$
is the centralized linear-combining vector and ${\bf r}^{\rm{ul}}[k]\in \mathbb{C}^{ML}$ is the collective
\gls{ul} data signal, given by
\begin{equation} \label{eq:received-data-central2}
{\bf r}^{\rm{ul}}[k] = \begin{bmatrix} {\bf r}_{1}^{\rm{ul}}[k] \\ \vdots \\  {\bf r}_{L}^{\rm{ul}}[k]
\end{bmatrix} = \sum_{i=1}^{K} {\bf h}_{i} x_i^{\rm{ul}}[k] +{\bf z}^{\rm ul}[k]
\end{equation}
with ${\bf z}^{\rm ul}[k]= [{\bf z}_{1}^{{\rm ul}^{\Ttran}}[k] \, \ldots \, {\bf z}_{L}^{{\rm
ul}^{\Ttran}}[k]]^{\Ttran} \in \mathbb{C}^{ML}$ being the collective noise vector.
Substituting~\eqref{eq:uplink-CPU-data-estimate} into~\eqref{eq:received-data-central2} we obtain
\begin{align}
\!\!y_i^{\rm{ul}}[k] 
= \underbrace{{\bf u}_{i}^{\Htran} {\bf h}_{i}}_{g} \underbrace{x_i^{\rm{ul}}[k]}_{q[k]} +
\underbrace{\sum_{j=1, j\ne i}^{K} {\bf u}_{i}^{\Htran} {\bf h}_{j} x_{j}^{\rm{ul}}[k] +{\bf u}_{i}^{\Htran}{\bf z}^{\rm ul}[k]}_{z[k]}\label{eq:uplink-CPU-data-estimate-new}
\end{align}
which can be expressed in the same form as~\eqref{eq:simplified_channel} if we set $v[k] =
y_{i}^{\mathrm{ul}}[k]$, $q[k] = x_{i}^{\mathrm{ul}}[k]$,
$g={\vecu}_{i}^{\Htran}{{\vech}}_{i}$, and $z[k] = \sum_{j=1,j\ne i}^{K}
{\vecu}_{i}^{\Htran}{\vech}_{j} x_{j}^{\mathrm{ul}}[k] + {\vecu}_{i}^{\Htran}{\vecz}^{\rm ul}[k]$. Given all
channels and combining vectors, the random variables $\{z[k] :
k=1,\ldots,\nul\}$ are conditionally \iid and
$z[k]\sim \jpg(0, \sigma^2)$ with $\sigma^2 = \sigma\sub{ul}^2\vecnorm{\vecu_{i}}^2 +
\rho^{\mathrm{ul}}\sum_{j=1, j\neq i}^K \abs{\herm{\vecu}_{i} \vech_{j}}^2$.

We assume that the \gls{cpu} treats the channel estimate $\widehat{\vech}_i$ as perfect and that the transmitted codeword is drawn from a codebook $\setC^\mathrm{ul}$. The estimated codeword $\widehat{\bf x}_i^{\mathrm{ul}}$ is thus obtained by performing mismatched \gls{snn} decoding with $\widehat{g}= \herm{{\bf u}}_i \widehat{\vech}_i$, i.e.,
\begin{equation}\label{eq:mismatched_snn_decoder-uplink}
  \widehat{\bf x}_i^{\mathrm{ul}}=\argmin_{\widetilde{\bf x}_i^{\mathrm{ul}} \in \setC^\mathrm{ul}} \vecnorm{{\bf y}_i^{\mathrm{ul}}-\widehat{g}\widetilde{\bf x}_i^{\mathrm{ul}}}^2
\end{equation}
with ${\bf y}_i^{\mathrm{ul}} = [y_i^{\mathrm{ul}}[1],\ldots,y_i^{\mathrm{ul}}[\nul]]^{\Ttran}$ and $\widetilde{\bf x}_i^{\mathrm{ul}} = [\widetilde{x}_i^{\mathrm{ul}}[1],\ldots, \widetilde{x}_i^{\mathrm{ul}}[\nul]]^{\Ttran}$.
An upper bound on the packet error probability then follows by applying~\eqref{eq:rcus_fading}.
It is important to note that, although we assumed Rayleigh fading and \gls{mmse} channel estimation, the
obtained bound is actually valid for any channel law and channel-estimation method, as well as any choice of
the spatial combiner ${\bf u}_{i}$. For a
detailed discussion on centralized combining schemes, we refer the interested reader to~\cite[Sec. 5.1.3, Sec.
5.1.4]{Demir21}.

\subsubsection{Distributed Operation} 
In a distributed cell-free network, the channel estimates are computed locally at
the APs and are used to obtain local estimates of UE data. Hence, unlike a fully centralized network, AP $l$
can only use its own local channel estimates for the design of the local combiner ${\bf u}_{il}$. 
The locally spatially-filtered signals at each \gls{ap} are then sent to the CPU, which performs detection. 
We assume that the received signal at the CPU is
the average of the locally filtered signals, i.e.,
\begin{align}\label{eq:uplink-CPU-L2}
  y_i^{\rm{ul}}[k] &= \sum\limits_{l=1}^L{\bf u}_{il}^{\Htran} {\bf r}_{l}^{\rm{ul}}[k].
  \end{align}
In a distributed network, the CPU does not have
knowledge of channel estimates and thus only the statistics can be utilized for data detection. 
 Specifically, we assume that $\widehat{\bf x}_i^{\mathrm{ul}}$ is obtained as in~\eqref{eq:mismatched_snn_decoder-uplink} but with 
 \begin{align}\label{eq:distr-hard}
\widehat{g} = \Ex{}{{\vecu}_{i}^{\Htran}{\vech}_{i}} = \sum_{l=1}^L\Ex{}{{\vecu}_{il}^{\Htran}{\vech}_{il}}.
\end{align}
As in the centralized case, the upper bound on the error probability, obtained from~\eqref{eq:rcus_fading}, is
valid for any fading-channel distribution  and any channel-estimation method as well as for any choice of local
combiners $\{{\bf u}_{il}: l=1,\ldots,L\}$. For a
detailed discussion on the choice of combiners, we refer the interested reader to~\cite[Sec.
6.1.2]{Demir21}.

\subsection{Downlink Operation}\label{sec:dlDataPhase}
We now consider the DL counterparts of the two UL operations described above.

\subsubsection{Centralized Operation}

In a centralized network, the CPU exploits channel reciprocity to obtain estimates of the collective channel
vectors, which are then used to compute the precoding vectors. 
We assume that
\begin{equation} 
{\bf w}_{i} = \sqrt{\rho_i^{\mathrm{dl}}}\bar {\bf w}_{i} 
\end{equation}
where $\vecnorm{ \bar{\bf w}_{i}}^2 = 1$ so that $\rho_i^{\mathrm{dl}}$ can be thought as the \gls{dl}
transmit power. Different precoders yield different tradeoffs between the error probability achievable at the
\glspl{ue}. A common heuristic comes from \gls{ul}-\gls{dl} duality~\cite[Sec.~4.3.2]{bjornson19}, which
suggests to choose the precoding vectors ${\bf w}_i$ as the following function of the combining vectors: ${\bf
w}_i= {{\bf u}_i}/{\sqrt{\Ex{}{\vecnorm{{\bf u}_i}^2}}}$.

As in UL, we can put~\eqref{eq:Cell-free-DL} in the same form as~\eqref{eq:simplified_channel} by setting
$v[k] = y_{i}^{\mathrm{dl}}[k]$, $q[k] = x_{i}^{\mathrm{dl}}[k]$, $g = {\vech}_{i}^{\Htran} {\vecw}_{i}$, and
$z[k] = \sum_{j=1,i\ne i}^{K}  {\vech}_{i}^{\Htran} {\vecw}_{j} x_{j}^{\mathrm{dl}}[k] +
z_{i}^{\mathrm{dl}}[k]$. 
The random variables $\{z[k] : k=1,\ldots,\ndl\}$ are \iid with $z[k]\sim
\jpg(0, \sigma^2)$ and $\sigma^2 = \sigma\sub{dl}^2 + \rho^{\mathrm{dl}}\sum_{i^{\prime}=1, i\neq
i^{\prime}}^K \abs{\herm{\vech_{i}}\vecw_{i^{\prime}}}^2$. 
The estimated codeword is thus obtained by performing mismatched SNN decoding as 
\begin{equation}\label{eq:mismatched_snn_decoder-donwlink}
  \widehat{\bf x}_i^{\mathrm{dl}}=\argmin_{\widetilde{\bf x}_i^{\mathrm{dl}} \in \setC^\mathrm{dl}} \vecnorm{{\bf y}_i^{\mathrm{dl}}-\widehat{g}\widetilde{\bf x}_i^{\mathrm{dl}}}^2
\end{equation}
with ${\bf y}_i^{\mathrm{dl}} = [y_i^{\mathrm{dl}}[1],\ldots,y_i^{\mathrm{dl}}[\ndl]]^{\Ttran}$ and $\widetilde{\bf x}_i^{\mathrm{dl}} = [\widetilde{x}_i^{\mathrm{dl}}[1],\ldots, \widetilde{x}_i^{\mathrm{dl}}[\ndl]]^{\Ttran}$. 

Note that without pilot transmission in the \gls{dl}, the \gls{ue} has no knowledge
of the precoded channel $\xi_{ii}= {\bf h}_i^{\Htran}{\bf w}_i$ in \eqref{eq:Cell-free-DL}. 
We assume, however, that the \gls{ue} is aware of its expected value $\Ex{}{{\bf h}_i^{\Htran} {\bf w}_i}$ and uses this quantity to perform mismatched \gls{snn} decoding. 
Specifically, we set $\widehat{g} = \Ex{}{{\bf h}_i^{\Htran} {\bf w}_i}$
in~\eqref{eq:mismatched_snn_decoder-donwlink}. When pilots are transmitted in
the \gls{dl}, mismatched SNN decoding is performed with $\widehat{g} = \widehat
\xi_{ii}$ where $\widehat \xi_{ii}$ is computed using, e.g., the LS estimator~\eqref{eq:ls_channel_est}.

\subsubsection{Distributed Operation} 
We assume that the CPU produces the downlink codewords $\{{\bf x}_i^{\mathrm{dl}}: i=1,\ldots,K\}$ and send them to the serving APs. 
Each AP then performs spatial precoding on the basis of the available local channel estimates. 
For example the signal transmitted by \gls{ap} $l$ in channel use $k$ is given by 
\begin{equation}
\sum_{i=1}^{K}  {\bf w}_{i l} x_{i}^{\rm dl}[k].
\end{equation}
%

As in the centralized case, UE $i$ detects the transmitted codeword by performing the mismatched SNN decoding
operation in~\eqref{eq:mismatched_snn_decoder-donwlink}. 
Specifically, if no DL pilots are transmitted, the \gls{ue} $i$ sets $\widehat{g} = \Ex{}{{\bf h}_i^{\Htran} {\bf w}_i}$. 
If DL pilots are transmitted, UE $i$ sets $\widehat{g} = \widehat \xi_{ii}$.

 
 \begin{table}[t]
	\renewcommand{\arraystretch}{1.}
	\centering
		\caption{Network parameters.}
	\begin{tabular}{|c|c|}
		\hline \bfseries $\!\!\!\!\!$ Parameter $\!\!\!\!\!$ & \bfseries Value\\
		\hline\hline
				Network area &  $150$\,m $ \times\,  150$\,m \\
				Number of UEs &  $K=40$ \\
				Bandwidth & $B = 20$\,MHz  \\
		
		Receiver noise power & $\sigma_{\mathrm{ul}}^2 = \sigma_{\mathrm{dl}}^2=-96$\,dBm \\
		
		Number of information bits & $b = \log_2 m = 160$ \\
		
		Total number of channel uses & $n = 300$ \\
		
		Number of UL/DL channel uses & $n/2 = 150$ \\
		
		Number of UL pilots & $\np^{\rm ul}=40$ \\
		
		Number of DL pilots & $\np^{\rm dl}=0$ or $40$ \\
		
		Distance between UE $i$ and AP $l$ & $d_{il}$\\
		
		Large scale fading $\beta_{il}$ in dB & $ -30.5 - 37.6\log_{10}\left( \frac{d_{il}}{1\,\text{m}}\right)$\\
		
		Height difference between AP and UE & $10$\,m \\
		
		Target error probability & $\epsilon\sub{target} = 10^{-5}$\\
		
		Combing/precoding scheme & MMSE or MR\\
		
		\hline
	\end{tabular}
	\label{tab:system_parameters_running_example}
\end{table}
\section{Numerical Analysis}\label{sec:numericalResult}
We present numerical simulations to characterize the UL and DL performance of the different cell-free Massive MIMO implementations in the \gls{urllc} regime. 
We consider an automated-factory propagation scenario with no wrap-around topology. The URLLC requirements for our numerical experiments, which are
provided in Table~\ref{tab:system_parameters_running_example}, have been selected according to the 3GPP technical
specifications~\cite{3GPP22.104}. 
The number of UL and DL channel uses corresponds roughly to a resource block in 5G NR and yields a latency of around
$100\mus$. 
When multiple antennas are used at the APs, the spatial correlation matrices are generated using the local scattering model from~\cite[Sec.~2.6]{bjornson19}. 
Specifically, we assume that the scatterers are uniformly distributed in the angular interval $[\varphi_{i}
-\Delta, \varphi_{i} + \Delta]$, where  $\varphi_{i}$ is the nominal angle-of-arrival of \gls{ue} $i$, where
$i=1,\dots,K$, and $\Delta$ is the angular spread. 
Hence, the $(m_1,m_2)$th element of ${\bf R}_{il}$ is equal to \cite[Sec.~2.6]{bjornson19}
\begin{align}\label{eq:2DChannelModel}
\left[ {\bf R}_{il} \right]_{m_1,m_2} =\frac{\beta_{il}}{2\Delta} \int_{-\Delta}^{\Delta}{ e^{\mathsf{j} \pi(m_1-m_2) \sin(\varphi_{i} + {\bar \varphi}) }}d{\bar \varphi}.
\end{align}
In all subsequent simulations, we assume $\Delta = 25^\circ$. 

The analysis is carried out by using the network availability $\eta$ defined in~\eqref{eq:network_avail} as performance metric. 
Specifically, for fixed \gls{ue} positions, we compute the average \gls{ul} and \gls{dl} error probabilities
$\epsilon^{\mathrm{ul}}$ and $\epsilon^{\mathrm{dl}}$ for an arbitrary \gls{ue} within the coverage area by
averaging over the small-scale fading and the additive noise.
Then, we evaluate the probability, computed with respect to the random user positions, that $\epsilon^{\mathrm{ul}}$ or
$\epsilon^{\mathrm{dl}}$ are below $\epsilon\sub{target}=10^{-5}$.
Similar to Section~\ref{sec:benefit_cell_free}, the use of \gls{mmse} combining and
precoding in both the UL and the DL turns out to be mandatory to achieve high $\eta$. 
Hence, we will focus on \gls{mmse} combining and precoding in the reminder of
the section. 

When considering network architectures involving multiple APs, we assume that
the APs are located on a square grid within the coverage area. 
This implies that the number $L$ of APs is chosen so that $\sqrt{L}$ is an integer.
When considering cell-free architectures, we will also assume that $L\geq 16$.

To perform a fair comparison, we use the same propagation model for cellular and
cell-free simulations. 
Furthermore, the same UE locations and pilot assignments are used in both scenarios.
We assume that the cellular BSs are equipped with half-wavelength-spaced uniform linear arrays. 
The spatial correlation follows the model in~\eqref{eq:2DChannelModel}.
\subsection{Uplink}
We assume that  orthogonal pilot sequences of length $\np^{\rm ul}=40$ are used
in the \gls{ul} for \gls{mmse} channel estimation. Since $K=40$, pilot
contamination is avoided. In the centralized case, we use the MMSE combiner~\cite{bjornson20-1a}
\begin{align}
{\bf u}_{i} = \rho^{\mathrm{ul}} \left(\sum_{i' =1 }^K \rho^{\mathrm{ul}}\left(\widehat \vech_{i} \widehat \vech_{i} ^{\Htran} + {\bf C}_{i} \right) + \sigma\sub{ul}^2{\bf I}_{LM}\right)^{-1}\widehat \vech_{i}
\end{align}
while the local  MMSE combiner
\begin{align}\label{eq:local-mmse}
{\bf u}_{il} = \rho^{\mathrm{ul}} \left(\sum_{i' =1 }^K \rho^{\mathrm{ul}}\left(\widehat \vech_{il} \widehat \vech_{il} ^{\Htran} + {\bf C}_{il} \right) + \sigma\sub{ul}^2{\bf I}_{M}\right)^{-1}\widehat \vech_{il}
\end{align}
is used with decentralized operation.
In~\eqref{eq:local-mmse}, ${\bf C}_{il} = {\bf R}_{il} - {\bf \Phi}_{il}$ is the covariance matrix of the estimation
error $\widetilde {\bf h}_{il} = {\bf h}_{il} - \widehat {\bf h}_{il}$.
MMSE channel estimation is used with both architectures.
  \begin{figure}[t!]
	\centering 
	\begin{overpic}[width=0.95\columnwidth,tics=10]{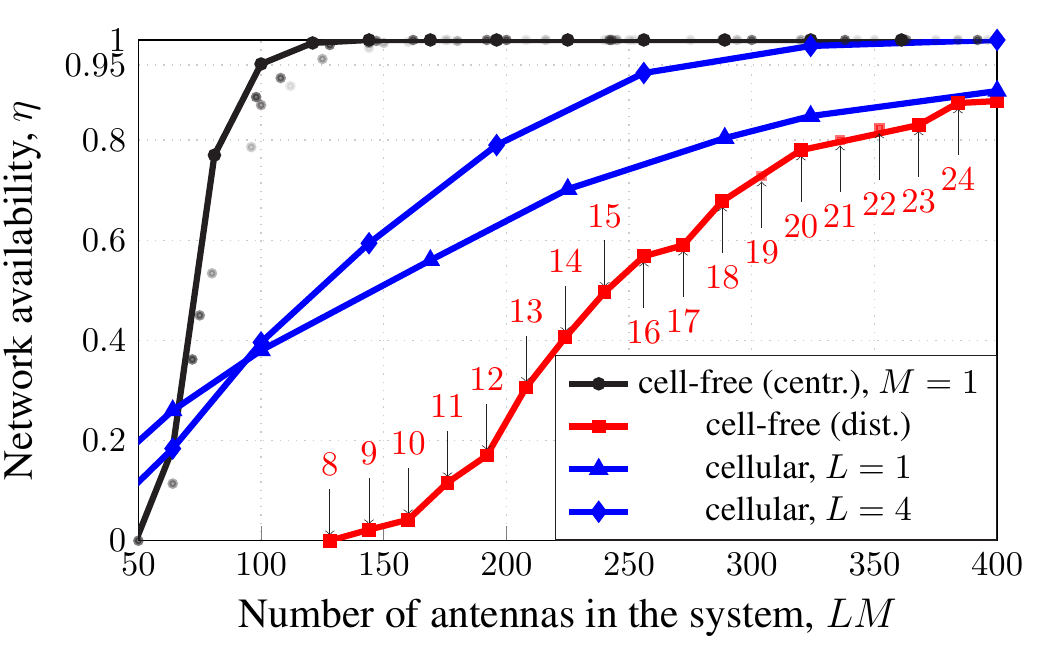}
	\end{overpic}  
	\caption{\gls{ul} network availability $\eta$ with different network deployments as a
    function of $LM$. The labels indicate the number of AP antennas $M$
that result in the largest $\eta$ for the cell-free distributed
case.} 
	\label{fig:section4_figure4} 
\end{figure}

\subsubsection{Network availability vs. number of
antennas}\label{sec:ul-eta-ant}
In Fig.~\ref{fig:section4_figure4}, we plot the network availability for
$\rho^{\mathrm{ul}}=-10$ dBm,  as a
function of the total number of antennas $LM \in\{50,\dots, 400\}$.
We see from Fig.~\ref{fig:section4_figure4} that the centralized cell-free architectures allows one
to achieve a network availability $\eta\geq 0.95$ when $LM\geq 100$. 
Using single-antenna APs ($M=1$) is optimal in the centralized case: the
network availability 
obtainable for larger values of $M$ is
strictly smaller.\footnote{
The network availability 
obtainable in the centralized cell-free setting when $M\in \{2,\dots,15\}$ is indicated in
Fig.~\ref{fig:section4_figure4} by the black dots, which are color-coded from
darker ($M=2$) to lighter  ($M=15$). The same convention is used also in
Fig.~\ref{fig:net_avail_MMSE}.}
This implies that, when providing URLLC services, it is advantageous to
distribute as much as possible the available antennas over the coverage area to reduce the average
distance between APs and UEs.
A cellular architecture with $L=4$ BSs requires at least $M=64$ antennas per
BS (i.e., $LM=256$) to achieve $\eta\geq 0.95$. 
This value of network availability cannot be achieved when $L=1$, even when
$M=400$, confirming again the importance of reducing the average distance
between BSs and UEs. 
For the values of $LM$ considered in the figure, the distributed cell-free
architecture, whose performance is optimized over
$M\in\{8,\dots,25\}$, does not achieve $\eta\geq 0.95$. 
Note that, in this architecture, $M=1$ is not optimal. 
On the contrary, the number of antennas $M$ per APs (indicated in the figure by the
red labels) needs to be increased as $LM$ is increased to maximize $\eta$. 
This comes as no surprise since~\eqref{eq:distr-hard} provides an accurate
channel estimate 
only when channel hardening occurs, which requires the APs to have sufficiently
many antennas. 

\begin{figure}[t!]     
  \begin{subfigure}{\columnwidth}
    \centering
    \vspace{0.45cm}
    \includegraphics[width=0.95\columnwidth]{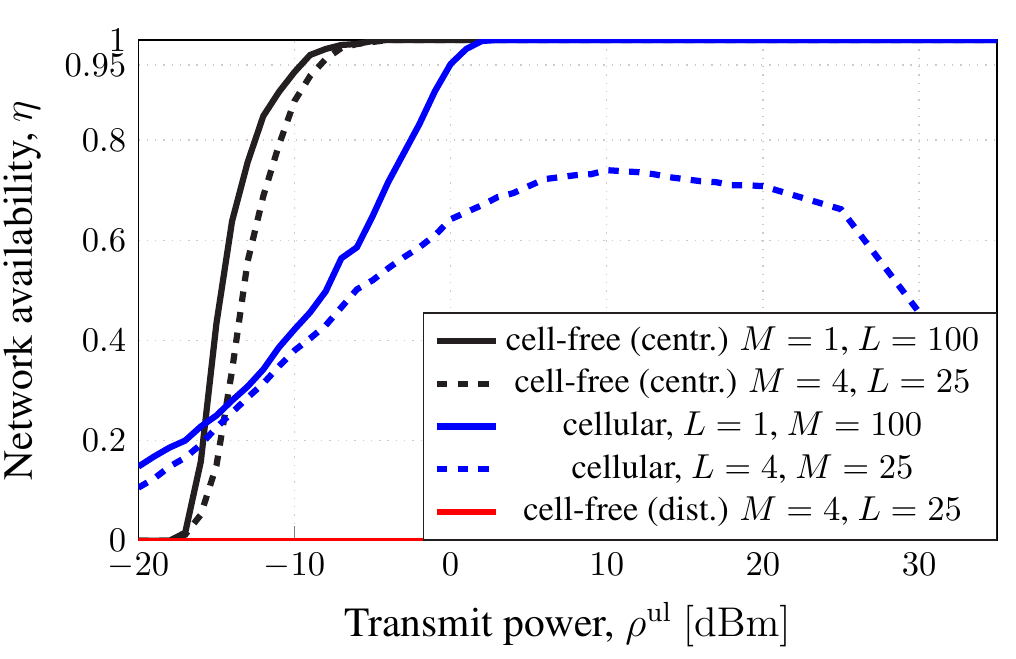}
    \caption{$LM = 100$}
    \label{fig:rho_vs_netavail_UL_MMMSE_LM_100}
  \end{subfigure}
  
    \begin{subfigure}{\columnwidth}
    \centering
    \includegraphics[width=0.95\columnwidth]{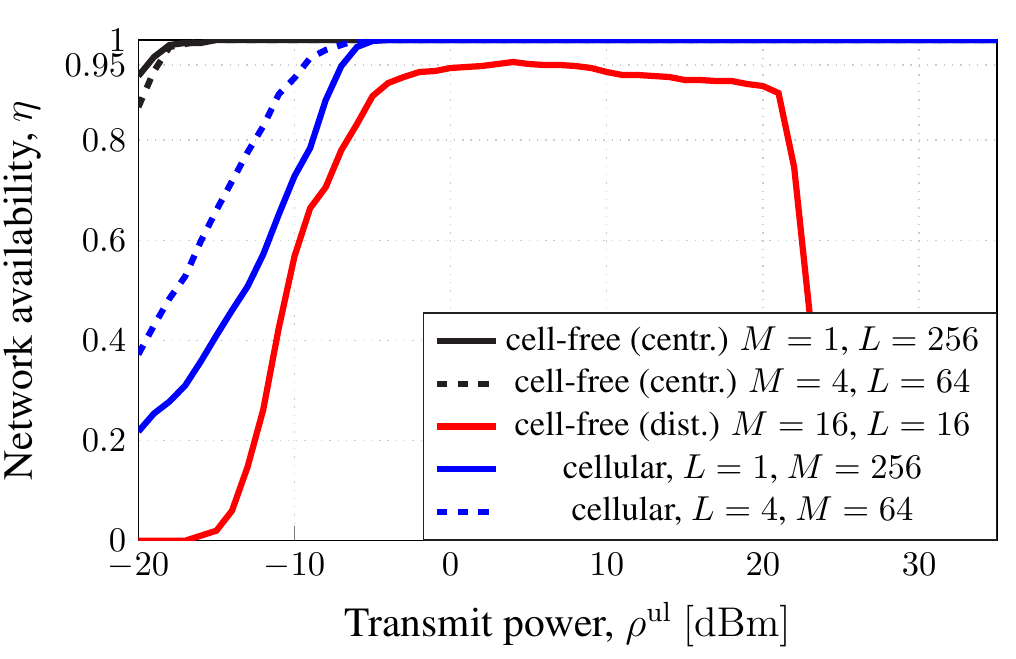}
   \caption{$LM = 256$}
    \label{fig:rho_vs_netavail_UL_MMMSE_LM_256}
  \end{subfigure}
         \caption{\gls{ul} network availability as a function of the transmit power for $LM=100 $ and $256$.}\vspace{-0.5cm}
       \label{fig:section4_figure5}
  \end{figure}
\subsubsection{Network availability vs. transmit power}\label{sec:ul-eta-pow}
In Fig.~\ref{fig:section4_figure5}, we plot $\eta$ as a function of the 
UL transmit power $\rho^{\rm ul}$ for $LM \in \{100,256\}$.
For the case $LM=100$, we consider two centralized cell-free architectures, one
with $L=100$ single-antenna APs and one with $L=25$ APs with $4$ antennas, as well as
two cellular architectures, one with a single BS ($L=1$) and one with $L=4$
BSs. 
The results of Fig.~\ref{fig:section4_figure5} show that the distributed cell-free architecture does not support positive $\eta$. Indeed, the APs have too few antennas to make channel-hardening-based spatial processing work.
Furthermore, the centralized cell-free architecture with
single-antenna APs achieves $\eta\geq 0.95$ for 
 $\rho^{\rm ul} \ge -17$ dBm. 
The cellular architecture with $L=1$ requires $\rho \geq 0\dBm$, whereas the
cellular architecture with $L=4$ is not able to achieve $\eta\geq 0.95$,
since intercell interference causes a degradation of $\eta$ when $\rho^{\rm
ul}$ is increased beyond $10$ dBm.

For the case $LM=256$, the performance of both centralized cell-free and
cellular improve, as expected.
Unlike Fig.~\ref{fig:rho_vs_netavail_UL_MMMSE_LM_100}, the cellular network with
$L=4$ outperforms the one with $L=1$.
This means that $M=64$ antennas are sufficient at each of the $4$ BS to provide
accurate enough interference management, for the system to benefit from the
lower average UE-BS distance.
We also illustrate the performance of a distributed
cell-free network with $L=16$ \glspl{ap}: this architecture does not achieve
$\eta\geq 0.95$. 
\subsection{Downlink}
We analyze next the DL network availability, with and without DL pilots. 
When pilot sequences are transmitted, we assume that they are orthogonal and that
$\np^{\rm dl} = 40$, so that $\nul = \ndl = 110$.
Furthermore, we assume that the LS channel estimator~\eqref{eq:ls_channel_est} is used.
\subsubsection{The impact of DL pilots} \label{sec:num_hardening}
Before investigating the network availability, we study first the impact of
DL pilots for a simple scenario. 
Specifically, we assume that a single UE is located in
the center of the coverage area and analyze the downlink packet error
probability achieved by a
centralized network with single-antenna APs, with and without DL pilots. 
\gls{mr} precoding is used and the transmit
power is $\rho^{\rm dl} = -10\dBm$. 
\begin{figure}[t!]     
    \centering
    \includegraphics[width=0.95\columnwidth]{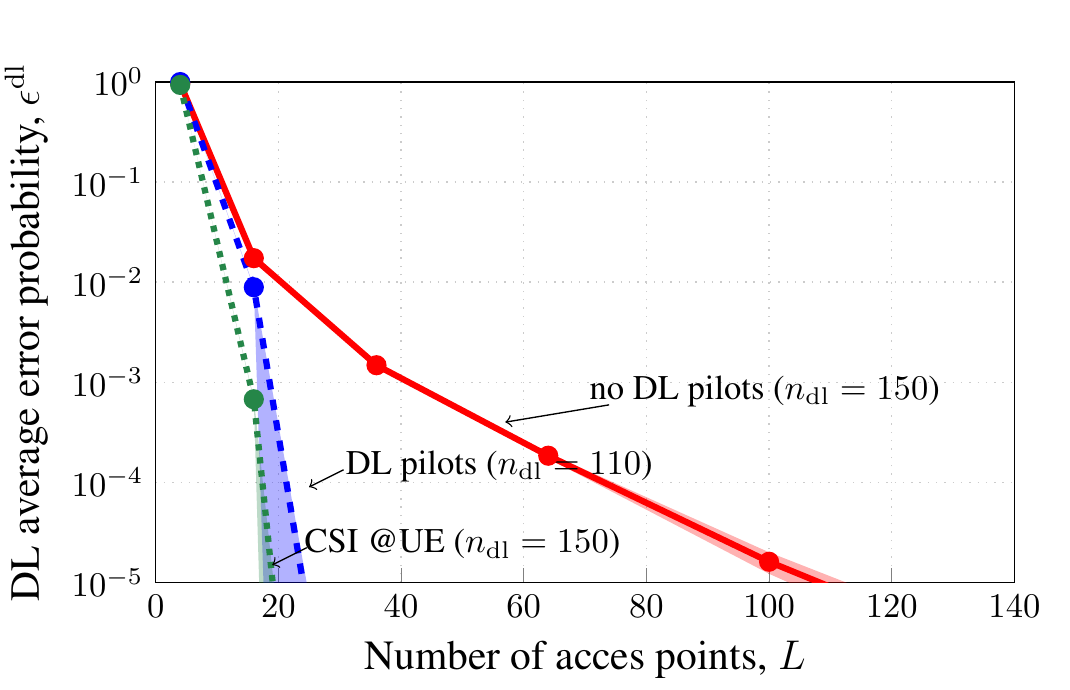}
    \caption{\gls{dl} average error probability $\epsilon^{\rm dl}$ for the
    centralized single-user single-antenna cell-free network as a function
    of the number of single-antenna access points $L$. \gls{mr} precoding is used for transmission.}
       \label{fig:hardening_study_cell-free_dens}
  \end{figure} 
Fig.~\ref{fig:hardening_study_cell-free_dens} shows the \gls{dl} average error
probability $\epsilon^{\rm dl}$ as a function of the number of APs $L$. 
The case where a genie provides
the UE with perfect knowledge of the precoded channel $g = {\bf h}^{\Htran} {\bf
w}$ is also reported as benchmark.  
The figure clearly illustrates that  DL pilots are beneficial in a centralized
network. 
Indeed, to achieve an average error probability
$\epsilon^{\rm dl}= 10^{-5}$, $L=25$ \glspl{ap} are sufficient.
To achieve the same average error probability without DL pilots, one needs 
$L=140$ \glspl{ap}. 
This means that the penalty incurred by reducing the available
channel uses for data transmission from $150$ to $110$ is much smaller than the benefit from having
an accurate estimate of the precoded channel $g = {\bf h}^{\Htran} {\bf w}$ at
the UE. 

\begin{figure}[t!]     
  \begin{subfigure}{\columnwidth}
    \centering
    \includegraphics[width=0.95\columnwidth]{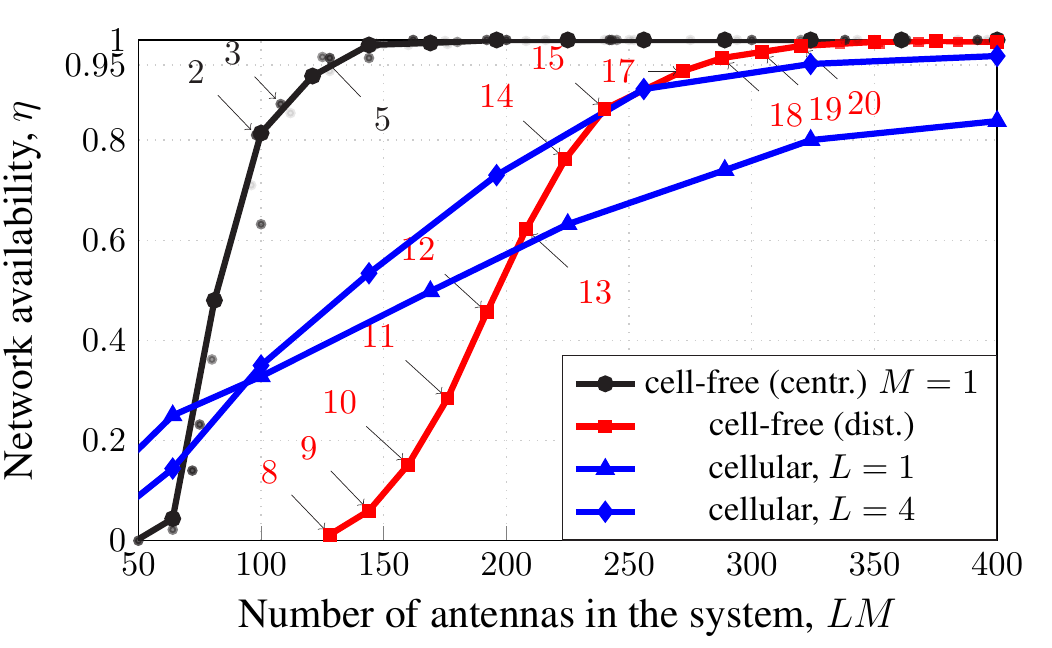}
    \caption{With \gls{dl} pilots}
    \label{fig:Msyst_vs_netavail_DL_MMMSE_pilots}
  \end{subfigure}  
  \begin{subfigure}{\columnwidth}
    \centering
    \includegraphics[width=0.95\columnwidth]{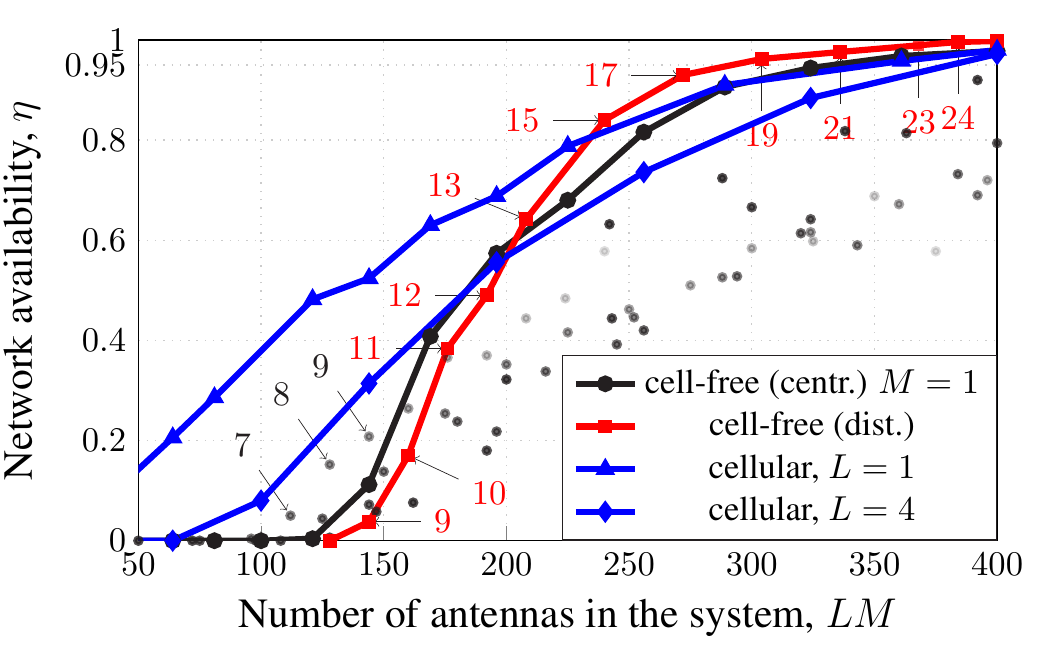}
    \caption{Without DL pilots}
    \label{fig:Msyst_vs_netavail_DL_MMMSE}
  \end{subfigure}
  \caption{\gls{dl} network availability for $\epsilon^{\rm dl}\sub{target} = 10^{-5}$ with \gls{mmse} precoding as a function of the total number of antennas in the system, $LM$. The labels indicate the number of AP antennas $M$ necessary to achieve the reported $\eta$.
  }
  \label{fig:net_avail_MMSE}
 \end{figure} 
\subsubsection{Network availability vs. number of antennas}\label{sec:net_avail_analysis}

 \begin{figure}[t!]     
   \begin{subfigure}{\columnwidth}
    \centering
    \includegraphics[width=0.95\columnwidth]{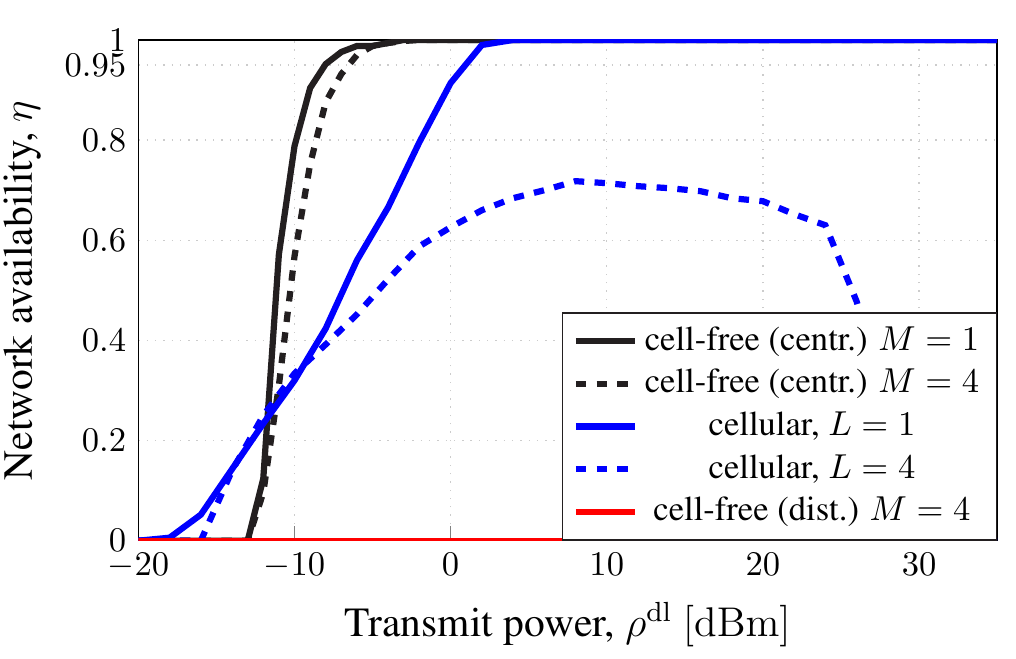}
    \caption{With \gls{dl} pilots}
    \label{fig:rho_vs_netavail_DLpilots_MMMSE_LM_100}
  \end{subfigure}
  \begin{subfigure}{\columnwidth}
    \centering
    \includegraphics[width=0.95\columnwidth]{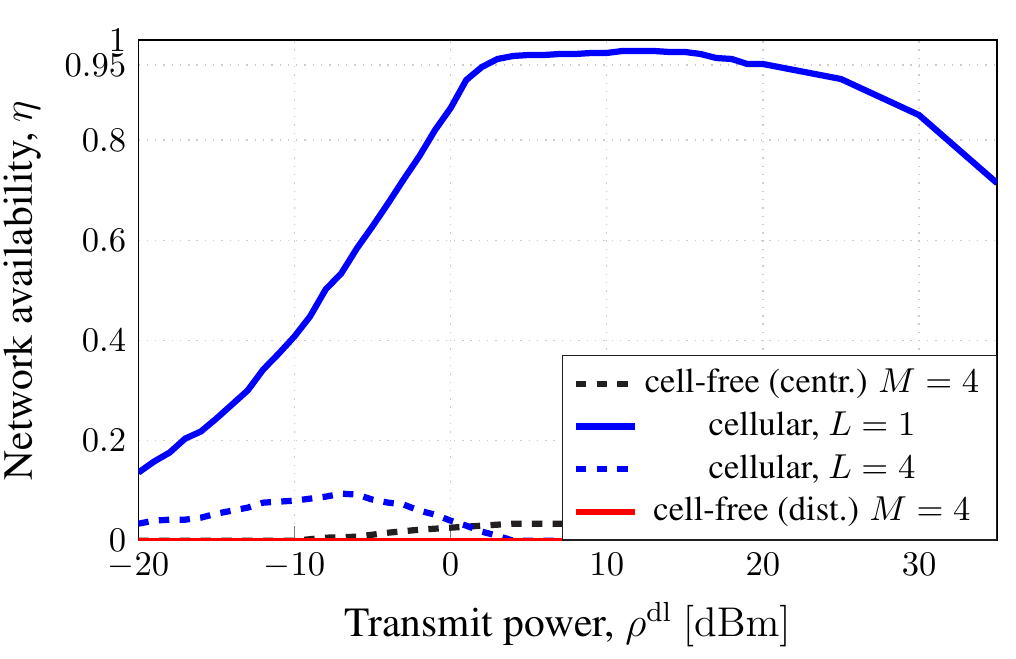}
    \caption{Without DL pilots}
    \label{fig:rho_vs_netavail_DL_MMMSE_LM_100}
  \end{subfigure}
  \caption{\gls{dl} network availability for $\epsilon^{\rm dl}\sub{target} = 10^{-5}$ with \gls{mmse} precoding as a function of the transmit power for $LM=100$.  
  }
  \label{fig:rho_net_avail_MMSE_100}
\end{figure} 
In Fig.~\ref{fig:Msyst_vs_netavail_DL_MMMSE_pilots}, we show $\eta$ as a
function of the total number $LM$ of antennas in the system, for the case of DL pilot transmission.
For the centralized cell-free and the cellular system, the observations one can draw from the figure are similar to the ones reported for the UL in
Section~\ref{sec:ul-eta-ant}: 
$M=1$ is
almost always optimal in the centralized cell-free case {(larger values of $M$, which are represented by the other black dots, yield worse performance in general)} and $L=4$ is superior to
$L=1$ in the cellular case.
The network availability achievable with the distributed cell-free system,
{optimized over $M\in \{8,\dots,25\}$}, is
larger than the one achievable in the UL, and this system achieves $\eta\geq
0.95$ when $LM\geq 270$. 
With the centralized cell-free architecture, $LM\geq 125$ is sufficient.

When no DL pilots are transmitted (see
Fig.~\ref{fig:Msyst_vs_netavail_DL_MMMSE}), the performance of the cellular network with
$L=1$ and of the distributed cell-free network actually improves slightly. 
Indeed, for
these two architectures, the use of DL pilots is not beneficial, because
the APs are equipped with a sufficiently large number of antennas for
channel-hardening-based estimates to be accurate.
So it is better to devote the channel uses spent on pilot symbols to data
transmission.
On the contrary, the performance of the centralized cell-free architecture
deteriorates significantly and is inferior to that of the distributed cell-free
system when $LM\geq 200$. 
\subsubsection{Network availability vs. transmit power}
Fig.~\ref{fig:rho_net_avail_MMSE_100} shows the \gls{dl}  network
availability with and
without DL pilots when $LM=100$, as a function of the DL transmit power. 
For this total number of antennas, the distributed cell-free architecture does
not achieve a positive network availability. 
When DL pilots are transmitted, the performance are similar as the one reported
for the UL in Section~\ref{sec:ul-eta-pow}: the centralized cell-free
with $M=1$ in the
most performing architecture, and the cellular architecture with $L=4$ suffers from
multi-cell interference for $\rho^{\rm dl}$ larger than around $8$ dBm.

When no DL pilots are transmitted, only the cellular architecture with $L=1$ is
capable of achieving $\eta\geq 0.95$. Centralized cell-free with $M=4$
(which outperforms the $M=1$ centralized architecture) and cellular
with $L=4$ yield $\eta \leq 0.1$.

\subsubsection{Do infinite-blocklength metrics provide accurate performance estimates?}
\begin{figure}[t] 
  \centering    
  \includegraphics[width=0.95\columnwidth]{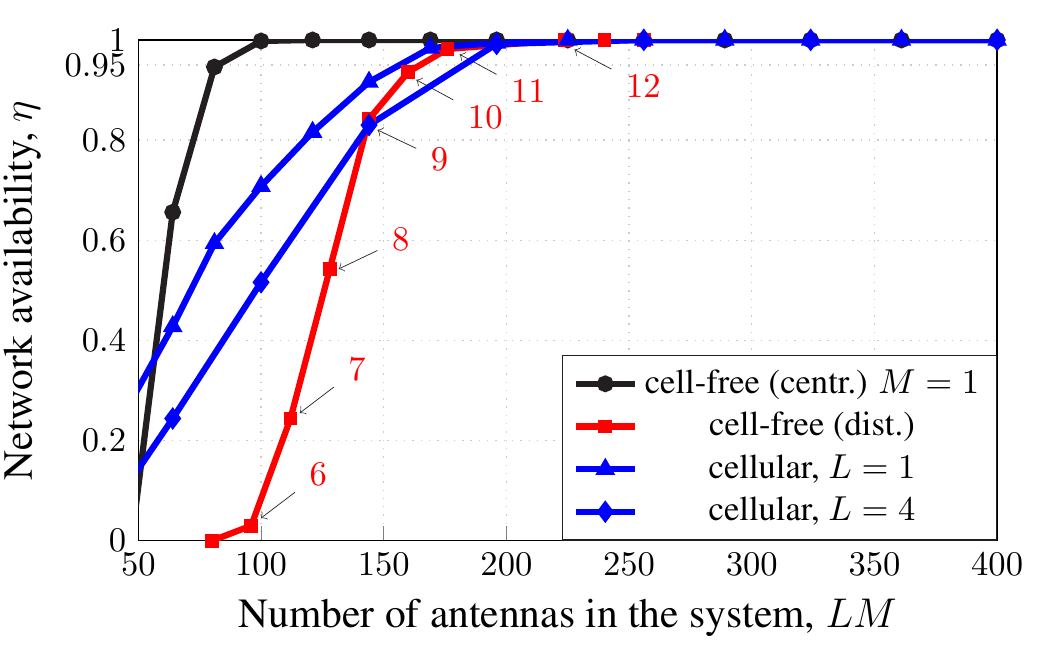}
  \caption{DL network availability $\eta$ computed using outage probability with different networks as a
  function of $LM$. The labels indicate the number of AP antennas $M$
that result in the largest $\eta$ for the cell-free distributed
case.}
       \label{fig:LM_comp_DL_out}
\end{figure}
In Fig.~\ref{fig:LM_comp_DL_out}, we depict the \gls{dl} network availability computed using as performance metric the outage probability: 
\begin{equation}
  \prob{\log\left(1+\frac{\snr{g}^2}{\sigma^2}\right)<R}.
\end{equation}
The setup is the same as the one considered in Fig.~\ref{fig:Msyst_vs_netavail_DL_MMMSE_pilots}.
By comparing Fig.~\ref{fig:Msyst_vs_netavail_DL_MMMSE_pilots} and Fig.~\ref{fig:LM_comp_DL_out}, we notice that outage
probability analyses yield overly optimistic results that can lead to misleading insights. 
One can show that similar conclusions hold also for the \gls{ul}.


\section{Conclusions}\label{sec:conclusions} 
We analyzed the performance of cell-free Massive
\gls{mimo} networks supporting the transmission of short packets under the high
reliability targets demanded in URLLC. The analysis 
was carried out using an accurate and
easy to evaluate approximation on the per-user \gls{ul}
and \gls{dl}  packet error error probability. 
This approximation relies on the saddlepoint expansion of a finite
blocklength bound (see Theorem~\ref{thm:rcus}). We showed that the saddlepoint approximation provided in Theorem~\ref{thm:saddlepoint} applies to both cellular and cell-free Massive MIMO networks, thereby generalizing the results by \"Ostman et al.~\cite{ostman20-09b}. Hence, while in the asymptotic regime of infinite blocklength, lower
bounds on the ergodic capacity are the primary tools to investigate numerically
the performance of both cellular and cell-free Massive MIMO systems and derive
insights into their design, in this
paper we show that, in the short-packet regime, one promising tool to achieve
the same goals is the
saddlepoint approximation on the RCUs error probability bound~\eqref{eq:simplified_channel}, computed for
the scaled nearest-neighbor decoding rule~\eqref{eq:mismatched_snn_decoder}. We used the packet size (blocklength) as a proxy for the latency in the access part of the network. 
We did not consider the contribution to the latency resulting from processing
delay or from transmission of information data over the fronthaul connecting central-processing unit and access point. This is an important issue that is left for future work. It turned out that, in a typical
automated-factory scenario, cell-free Massive \gls{mimo} with fully centralized
processing and single-antenna APs outperforms cell-free Massive \gls{mimo} with distributed
processing and cellular Massive \gls{mimo}.
However, for the centralized cell-free Massive \gls{mimo} architecture to perform
satisfactorily: 
\begin{inparaenum}[(i)]
    \item one must use \gls{mmse} linear processing. Indeed, \gls{mr} processing is not able to
        guarantee the reliability required in URLLC. 
    \item Furthermore, the APs need to transmit precoded pilot sequences in the DL, to allow the UEs to
        acquire a sufficiently accurate channel estimate. If DL pilots are not
        transmitted, one has to equip each AP with sufficiently many antennas to
        induce channel hardening. However, for a fixed total number of
        antennas per coverage area, this yields a reduction in the AP density,
        which affects performance negatively, because of the larger average distance
        between UEs and APs.
\end{inparaenum}

\bibliographystyle{IEEEtran}

\end{document}